\documentclass[12pt]{article}
\usepackage[letterpaper,top=2cm,bottom=2cm,left=2cm,right=2cm,marginparwidth=1.75cm]{geometry}
\usepackage{amsmath,amssymb}
\usepackage{graphicx}
\usepackage[colorlinks=true, allcolors=blue]{hyperref}
\usepackage{authblk}
\usepackage{tabularray}
\usepackage{bbm}
\usepackage{multirow}
\usepackage{cite}
\UseTblrLibrary{booktabs}

\def\mathswitchr#1{\relax\ifmmode{\mathrm{#1}}\else$\mathrm{#1}$\fi}
\def\mathswitch#1{\relax\ifmmode#1\else$#1$\fi}
\newcommand{\brc}[1]{\left(#1\right)}

\newcommand{\corr}{\mathswitchr{Corr}}
\newcommand{\bbeta}{\mathswitch{\boldsymbol{\beta}}}

\providecommand{\keywords}[1]
{
  \small
  \noindent
  \textbf{Keywords:} #1
}

\title{\textbf{\Large Goodness-of-fit Tests for Combined Unilateral and Bilateral Data}}
\author{Jia Zhou \footnote{\href{mailto:jiazhou@buffalo.edu}{jiazhou@buffalo.edu}} }
\author{Chang-Xing Ma \footnote{\href{mailto:cxma@buffalo.edu}{cxma@buffalo.edu}}}
\affil{Department of Biostatistics, University at Buffalo, Buffalo, NY 14214, USA}
\date{}

\begin{document}
\maketitle

\begin{abstract}
\fontsize{12pt}{14pt}\selectfont

\noindent
Clinical trials involving paired organs often yield a mixture of unilateral and bilateral data, where each subject may contribute either one or two responses under certain circumstances. While unilateral responses from different individuals can be treated as independent, bilateral responses from the same individual are likely correlated. Various statistical methods have been developed to account for this intra-subject correlation in the bilateral data, and in practice it is crucial to select an appropriate model for accurate inference. Tang \textit{et. al.} \cite{tang2012goodness} discussed model selection issues using a variety of goodness-of-fit test statistics for correlated bilateral data for two groups, and Liu and Ma \cite{liu2020goodness} extended these methods to settings with $g\ge2$ groups. 

In this work, we investigate the goodness-of-fit statistics for the combined unilateral and bilateral data under different statistical models that address the intra-subject correlation, including the Clayton copula model, in addition to those considered in prior studies. Simulation results indicate that the performance of the goodness-of-fit tests is model-dependent, especially when the sample size is small and/or the intra-subject correlation is high, which is consistent with the findings in \cite{liu2020goodness} for purely bilateral data. Applications to real data from otolaryngologic and ophthalmologic studies are included. \\

\keywords{
  Combined unilateral and bilateral outcomes;
  Clayton copula model;
  bootstrap procedures;
  Akaike information criterion;
  model selection techniques
}
\end{abstract}

\section{Introduction}
\label{sec:intro}
In clinical trial studies, the outcomes of paired organs, such as eyes, ears, and kidneys, are often recorded as binary data. The data are considered bilateral when both sites of the same individual are included, in contrast to unilateral data, where only one site of a paired organ is recorded per individual. Unilateral outcomes are typically assumed to be independent, with each individual’s response(s) being independent of others. Bilateral outcomes, however, exhibit intra-subject correlation due to the paired nature of the data. Both types of data are often recorded together in randomized clinical trials. For instance, in ophthalmologic studies, the focus is often on the statistical analysis of eyes rather than individuals. For $n$ enrolled patients, the medical records may include responses for between $n$ and $2n$ eyes, as some patients provide data for both eyes, while others provide data for only one. 

A number of statistical models has been proposed to address the intra-subject correlation problem. 
Rosner \cite{Rosner_1982} proposed a ``constant $R$ model" that the conditional probability of having a response in one eye, given a response in the other, is proportional to the marginal probability by a factor $R$, which coincides with the odds ratio. 
Dallal \cite{Dallal_1988} later argued that Rosner's constant $R$ model could lead a poor fit if the characteristic is almost certain to occur bilaterally with largely varying group specific prevalence. Instead, he proposed that the conditional probability is a constant $\gamma$. 
Subsequently, Donner \cite{Donner1989rhoModel} proposed an alternative approach that assumes a constant intra-person correlation $\rho$ for all the individuals in the sample. This model was proved robust with simulation study by Thompson \cite{Thompson_1993}. 
Clayton \cite{clayton1978model} proposed a model for association in bivariate variables using Clayton copula, which expresses the joint distribution in terms of the marginal cumulative distribution functions (CDFs) and a dependence parameter $\theta$. The Clayton copula model is particularly useful for capturing lower tail dependence, a clinically relevant feature in settings where disease in one organ increases the risk in the paired organ.
In addition to these parametric models, the independence model and saturated model are often used as benchmarks for describing bilateral and combined data structures. 

Different methods have been developed to analyze the correlated binary data under the aforementioned models, such as homogeneity test of proportions, and confidence interval estimation (e.g., see Refs. \cite{Tang_2008,Pei_2008,Pei_2011,pei2012confidence,Tang_2012,Pei_2014,Ma_2013R,liu2016testing,ma2017rho,shen2017testing,Zhuang_2018,Zhuang_2018CI,zhuang2018homogeneity,Peng_2019,shen2018CI,Shen_2019OR,Xue_2019,Xue_2019SMMR,lin2021fast,Shen2019reviewpaper,Yang_2019,li2020statistical,li2022statistical,li2023homogeneity,li2023testing,chen2022further,ma2021testing,mou2021homogeneity,mou2023homogeneity,mou2023asymptotic,ma2022testing,sun2022homogeneity,sun2022risk,zhang2023testing,hua2024testing,hua2024common,tian2024testing,wang2024homogeneity,wang2024interval,zhang2024simultaneous,liang2024homogeneity,liang2024exactrho,liang2024many,liang2025testing}). Given the variety of available models, it is essential to select an appropriate one for the data at hand. Previous work by Tang \textit{et. al.} \cite{tang2012goodness} and Liu and Ma \cite{liu2020goodness} investigated goodness-of-fit test methods for correlated bilateral data in the context of two and $g\ge2$ groups, respectively. In this paper, we focus on performing goodness-of-fit tests for the combined unilateral and bilateral data under the aforementioned models. Specifically, we compare the following six methods: deviance ($G^2$), Pearson Chi-square ($X^2$), adjusted chi-square ($X^2_{adj}$), along with three bootstrap methods ($B_1,B_2,B_3$). 

The rest of the paper is organized as follows. Section \ref{sec:models} introduces six models for analyzing combined binary data and describes the procedures for obtaining maximum likelihood estimates. Section \ref{sec:methods} presents the six methods for goodness-of-fit test. A simulation study is conducted in Section \ref{sec:simulation} to evaluate the performance of these methods in terms of empirical type I error rates and powers under different models. In Section \ref{sec:real-data}, three real-world examples are applied to illustrate the goodness-of-fit test methods. Conclusions are provided in Section \ref{sec:conclusions}.

\section{Models for Combined Unilateral and Bilateral Data}
\label{sec:models}
Let $m_{ri}$ be the number of individuals who contribute data on paired organs with $r~\brc{r=0,1,2}$ responses \footnote{The response means the organ being cured/affected.} in the $i$-th group ($i=1,\ldots,g$), and $n_{ri}$ be the number of individuals who contribute data on one of the paired organs with $r~\brc{r=0,1}$ responses in the $i$-th group, respectively. Let $m_{r+}~\brc{r=0,1,2}$ and $n_{r+}~\brc{r=0,1}$ be the numbers of subjects who respectively contribute to bilateral data and unilateral data with $r$ responses. Then, 
$$
m_{r+}=\sum_{i=1}^gm_{ri}, 
\quad
n_{r+}=\sum_{i=1}^gn_{ri}. 
$$

Similarly, let $m_{+i}$ and $n_{+i}$ be the numbers of subjects who contribute on unilateral and bilateral data in the $i$th group, respectively. Then, 
$$
m_{+i}=\sum_{r=0}^2m_{ri}, 
\quad
n_{+i}=\sum_{r=0}^1n_{ri}. 
$$
The total number of subjects in the study thus is 
$$
\sum_{i=1}^g\brc{\sum_{r=0}^2m_{ri}+\sum_{r=0}^1n_{ri}}=\sum_{i=1}^g\brc{m_{+i}+n_{+i}}=m_{++}+n_{++}. 
$$
The data structure is demonstrated in Table \ref{tab:data_struc}.
\begin{table}[thpb]
    \centering
    \caption{Frequency table for number of cured organs for subjects in $g$ groups.}
    \label{tab:data_struc}
    \begin{tabular}{cccccc}
    \toprule
     &\multicolumn{4}{c}{group} & \\
    \cline{2-5}
    \# of cured organs &1 &2 &$\ldots$ &g &total \\
    \midrule
    0 &$m_{01}$ &$m_{02}$ &$\ldots$ &$m_{0g}$ &$m_{0+}$ \\
    1 &$m_{11}$ &$m_{12}$ &$\ldots$ &$m_{1g}$ &$m_{1+}$ \\
    2 &$m_{21}$ &$m_{22}$ &$\ldots$ &$m_{2g}$ &$m_{2+}$ \\
    total &$m_{+1}$ &$m_{+2}$ &$\ldots$ &$m_{+g}$ &$m_{++}$ \\
    \midrule
    0 &$n_{01}$ &$n_{02}$ &$\ldots$ &$n_{0g}$ &$n_{0+}$ \\
    1 &$n_{11}$ &$n_{12}$ &$\ldots$ &$n_{1g}$ &$n_{1+}$ \\
    total &$n_{+1}$ &$n_{+2}$ &$\ldots$ &$n_{+g}$ &$n_{++}$ \\
    \bottomrule
    \end{tabular}
\end{table}

It is obvious that for the $i$-th group, the random variables $\brc{m_{0i},m_{1i},m_{2i}}$ and $\brc{n_{0i},n_{1i}}$ follow multinomial distributions. In particular,
  \begin{equation}
  \begin{aligned}
      &\brc{m_{0i},m_{1i},m_{2i}}\sim Multinomial\brc{m_{+i},p_{0i},p_{1i},p_{2i}}, \quad
      \sum_{r=1}^2p_{ri}=1, \\
      &n_{1i} \sim Binomial\brc{n_{+i},\pi_i}, 
  \end{aligned}  
  \label{eq:distribution}
  \end{equation}
where $p_{ri}$ denotes the probability of having $r$ ($r=0,1,2$) responses for a subject in the $i$-th group for bilateral data, and $\pi_i$ is the marginal probability of the response for a subject in the $i$-th group. 
Let $Z_{ijk}=1$ be the response of the $k$-th ($k=1,2$) paired organ for the $j$-th subject in the $i$-th group, then the joint probabilities $p_{ri}$ ($r=0,1,2$) read
\begin{equation}
    \begin{aligned}
        &p_{2i}=Pr\brc{Z_{ij1}=1,Z_{ij2}=1}=E\brc{Z_{ij1}Z_{ij2}}=\pi_i\left[\pi_i+\brc{1-\pi_i}\corr\brc{Z_{ij1},Z_{ij2}}\right], \\
        &p_{1i}=\sum_{k=1}^2Pr\brc{Z_{ijk}=1,Z_{ij,3-k}=0}=2\brc{\pi_i-p_{2i}}=2\pi_i\brc{1-\pi_i}\left[1-\corr\brc{Z_{ij1},Z_{ij2}}\right], \\
        &p_{0i}=Pr\brc{Z_{ij1}=0,Z_{ij2}=0}=1-p_{1i}-p_{2i}=\brc{1-\pi_i}\left[1-\pi_i+\pi_i~\corr\brc{Z_{ij1},Z_{ij2}}\right],   
    \end{aligned}
    \label{eq:prob_ri}
\end{equation}
where $\corr\brc{Z_{ij1},Z_{ij2}}$ denotes the intra-subject correlation between the two responses from the $j$-th subject in the $i$-th group. 

Various statistical models that account for within-subject dependence often introduce additional (nuisance) parameters to capture the intra-subject correlation. In what follows, we consider four such parametric models that introduce one nuisance parameter : i) Rosner's model, ii) Donner's (constant $\rho$) model, iii) Dallal's model, and iv) Clayton copula model, respectively. Let the nuisance parameter be $\kappa$ for the moment, then the log-likelihood function depends on $\bbeta=\brc{\pi_1,\ldots,\pi_g,\kappa}$, i.e.,
\begin{equation}
    l\brc{\pi_1,\ldots,\pi_g,\kappa|\brc{m,n}}=\sum_{i=1}^g\sum_{r=0}^2m_{ri}\log\brc{p_{ri}}+\sum_{i=1}^g\left[n_{0i}\log\brc{1-\pi_i}+n_{1i}\log\brc{\pi_i}\right]+\text{const},
    \label{eq:ll:generic}
\end{equation}
where dependence of $\kappa$ is through the joint probabilities $p_{ri}$ in (\ref{eq:prob_ri}), $\brc{m,n}$ denotes a vector of combined bilateral and unilateral data that
\begin{equation}
\brc{m,n}=\brc{m_{01},m_{11},m_{2,1},\ldots,m_{0g},m_{1g},m_{2g},n_{01},n_{11},\ldots,n_{0g},n_{1g}}, 
\label{eq:sample}
\end{equation}
and `const" is a constant term depending on $\brc{m,n}$. 

Suppose the log-likelihood is concave and certain regularity conditions are satisfied. The maximum likelihood estimates (MLEs) of $\bbeta$ can be obtained via the following normal equations
\begin{equation}
        \frac{\partial l\brc{\bbeta}}{\partial\pi_i}=0, \quad i=1,\ldots,g\,; \quad
        \frac{\partial l\brc{\bbeta}}{\partial\kappa}=0. 
        \label{eq:norm}
\end{equation}

Usually there are no closed-form solutions for the above normal equations, rather, the MLEs $\hat{\bbeta}=\brc{\hat{\pi}_1,\ldots,\hat{\pi}_g,\hat{\kappa}}$ can be solved iteratively. The iteration procedure for estimating the MLEs of $\pi_i$ and $\kappa$ is outlined below. 
\begin{enumerate}
    \item At the $\brc{t}$-th step, for given $\hat{\kappa}^{(t)}$, obtain $\hat{\pi}_i^{(t)}$ as a real root for equation $\partial l\brc{\bbeta}/\partial\pi_i=0$ for $i=1,\ldots,g$, such that $\hat{\bbeta}^{(t)}=\brc{\hat{\pi}_i^{(t)},\ldots,\hat{\pi}_g^{(t)},\hat{\kappa}^{(t)}}$.
    
    \item At the $\brc{t+1}$-th step, $\hat{\kappa}^{(t+1)}$ is evaluated with the Newton-Raphson method: 
    $$
    \hat{\kappa}^{(t+1)}=\hat{\kappa}^{(t)}+\brc{\left.-\frac{\partial^2l\brc{\bbeta}}{\partial\kappa^2}\right|_{\bbeta=\hat{\bbeta}^{(t)}}}^{-1}\left.\frac{\partial l\brc{\bbeta}}{\partial\kappa}\right|_{\bbeta=\hat{\bbeta}^{(t)}}.
    $$

    \item Repeat \textbf{Steps 1 - 2} until convergence of $\hat{\kappa}$ occurs, which can be measured by $\delta^{(t)}=\left|\hat{\kappa}^{(t+1)}-\hat{\kappa}^{(t)}\right|$. The iteration procedure stops when $\delta^{(t)}<\delta_0$ for a sufficiently small $\delta_0$, such as $10^{-6}$. 
\end{enumerate}

The initial values $\hat{\pi}_i^{(0)}$ and $\hat{\kappa}^{(0)}$ can be set somewhat arbitrarily but within their allowed regions. The region for $\kappa$ will be discussed in each model subsequently.

\subsection{Rosner's Model}
Rosner \cite{Rosner_1982} proposed a “constant $R$ model” that assumed equal dependence between two eyes of the same person for the phthalmologic data. More specifically, it assumed that the probability of cured eye at one site given cured eye at the other site for the $j$-th subject in the $i$-th group is proportional to the prevalence rate for the $i$-th group by a constant factor $R$, i.e., 
\begin{equation}
    Pr\brc{Z_{ijk}=1|Z_{ij,3-k}=1}=R\cdot Pr\brc{Z_{ijk}=1}=R\pi_i, 
\end{equation}
for $k=1,2$ denoting the left and right eye, respectively. The intra-subject correlation then reads $\corr\brc{Z_{ijk},Z_{ij,3-k}}=\brc{R-1}\pi_i/\brc{1-\pi_i}$. The region of $R$ is bounded by the region of probabilities and correlation. It can be shown that $R$ satisfies $0<R\le1/a$ if $a\le1/2$; $(2-1/a)/a\le R\le1/a$ if $a>1/2$ with $a=\underset{i=1,\ldots,g}{\max}\left\{\pi_i\right\}$ \cite{Ma_2013R}.

Taking $\kappa=R$ the normal equations in (\ref{eq:norm}) are derived as 
\begin{subequations}
    \begin{equation}
        \frac{\partial l\brc{\bbeta}}{\partial\pi_i}=
        \frac{2m_{0i}\brc{R\pi_i-1}}{1-2\pi_i+R\pi_i^2}+\frac{m_{1i}\brc{2R\pi_i-1}}{\pi_i\brc{1-R\pi_i}}+\frac{2m_{2i}}{\pi_i}-\frac{n_{0i}}{1-\pi_i}+\frac{n_{1i}}{\pi_i}=0, 
        \quad 
        \brc{i=1,\ldots,g}
        \label{eq:norm:mle1:rosner:1}
    \end{equation}
    \begin{equation}
        \frac{\partial l\brc{\bbeta}}{\partial R}=
        \sum_{i=1}^g\left[\frac{m_{0i}\pi_i^2}{1-2\pi_i+R\pi_i^2}+\frac{m_{1i}\pi_i}{R\pi_i-1}\right]+\frac{m_{2+}}{R}=0. 
        \label{eq:norm:mle1:rosner:2}
    \end{equation}
    \label{eq:norm:mle1:rosner}
\end{subequations}

The Eq. (\ref{eq:norm:mle1:rosner:1}) leads to a quartic equation w.r.t. $\pi_i$ as shown below, 
\begin{equation}
    a_i\pi_i^4+b_i\pi_i^3+c_i\pi_i^2+d_i\pi_i+e_i=0, 
    \label{eq:quartic:mle}
\end{equation}
where the coefficients are 
\begin{align*}
    &a_i=\brc{2m_{+i}+n_{+i}}R^2,  \\
    &b_i=-\brc{\brc{R+2}m_{0i}+\brc{2R+5}m_{1i}+2\brc{R+3}m_{2i}+3n_{0i}+\brc{R+3}n_{1i}}R, \\
    &c_i=\brc{4R+2}m_{0i}+\brc{7R+2}m_{1i}+4\brc{2R+1}m_{2i}+\brc{R+1}n_{0i}+2\brc{2R+1}n_{1i}, \\
    &d_i=-\brc{2\brc{R+1}m_{0i}+\brc{2R+3}m_{1i}+5m_{2i}+n_{0i}+\brc{R+3}n_{1i}}, \\
    &e_i=m_{1i}+2m_{2i}+n_i. 
\end{align*} 
Therefore, at the \textbf{Step 1} in the iteration procedure, $\hat{\pi}_i^{(t)}$ is a real root of the quartic equation (\ref{eq:quartic:mle}) for a given $\hat{R}^{(t)}$ \cite{ma2021testing}. The first (see the middle expression in the normal equation (\ref{eq:norm:mle1:rosner:2})) and second derivative of the log-likelihood w.r.t. $R$ 
\begin{equation}
\frac{\partial^2l\brc{\bbeta}}{\partial R^2}=-\sum_{i=1}^g\left[\frac{m_{0i}\pi_i^4}{\brc{R\pi_i^2-2\pi_i+1}^2}+\frac{m_{1i}\pi_i^2}{\brc{R\pi_i-1}^2}\right]+\frac{m_{2+}}{R^2}, 
\end{equation}
are used to update $\hat{R}^{(t+1)}$ in the $\brc{t+1}$-th iteration as described at \textbf{Step 2} in the iteration procedure.

\subsection{Donner's Model}
With Donner's approach \cite{Donner1989rhoModel}, the correlation between the outcomes of paired organs of a same subject is assumed to be the same in the sample such that 
\begin{equation}
    \corr\brc{Z_{ijk},Z_{ij,3-k}}=\rho, \quad
    \rho\in\left[-1,1\right]. 
    \label{eq:corr:rho}
\end{equation}
Taking $\kappa=\rho$, the normal equations have the following form 
\begin{subequations}
    \begin{equation}
        \frac{\partial l\brc{\bbeta}}{\partial\pi_i}=
        \frac{m_{0i}\brc{2\brc{1-\rho}\pi_i+\rho-2}}{\brc{1-\pi_i}\brc{1-\brc{1-\rho}\pi_i}}+\frac{m_{1i}\brc{1-2\pi_i}}{\pi_i\brc{1-\pi_i}}+\frac{m_{2i}\brc{2\brc{1-\rho}\pi_i+\rho}}{\pi_i\brc{\rho\brc{1-\pi_i}+\pi_i}}-\frac{n_{0i}}{1-\pi_i}+\frac{n_{1i}}{\pi_i}=0, 
        \label{eq:norm:mle1:donner:1}
    \end{equation}
    \begin{equation}
        \frac{\partial l\brc{\bbeta}}{\partial\rho}=
        \sum_{i=1}^g\left[\frac{m_{0i}\pi_i}{1-\brc{1-\rho}\pi_i}+\frac{m_{2i}\brc{1-\pi_i}}{\rho\brc{1-\pi_i}+\pi_i}\right]-\frac{m_{1+}}{1-\rho}=0, 
        \label{eq:norm:mle1:donner:2}
    \end{equation}
\end{subequations}
where a cubic equation
\begin{equation}
    a_i\pi_i^3+b_i\pi_i^2+c_i\pi_i+d_i=0, 
    \label{eq:cubic:pi:donner}
\end{equation}
can be derived from (\ref{eq:norm:mle1:donner:1}), and the respective coefficients read 
\begin{align*}
    &a_i=\brc{1-\rho}^2\brc{2m_{+i}+n_{+i}}, \\
    &b_i=\brc{1-\rho}\brc{\brc{3\rho-2}m_{0i}+3\brc{\rho-1}m_{1i}+\brc{3\rho-4}m_{2i}+\brc{\rho-1}n_{0i}+2\brc{\rho-1}n_{1i}}, \\
    &c_i=\rho\brc{\rho-2}m_{0i}+\brc{\brc{\rho\brc{\rho-4}+1}m_{1i}+\brc{\rho\brc{\rho-4}+2}m_{2i}+\brc{\rho\brc{\rho-3}+1}n_{1i}}, \\
    &d_i=\rho\brc{m_{1i}+m_{2i}+n_{1i}}. 
\end{align*}
Therefore, at the \textbf{Step 1} in the iteration procedure, $\hat{\pi}_i^{(t)}$ is a real root of the cubic equation (\ref{eq:cubic:pi:donner}) for a given $\hat{\rho}^{(t)}$ \cite{ma2022testing}. The first derivative shown in the middle expression in (\ref{eq:norm:mle1:donner:2}) and the second derivative of the log-likelihood w.r.t. $\rho$ 
\begin{equation}
    \frac{\partial^2l\brc{\bbeta}}{\partial\rho^2}=-\sum_{i=1}^g\left[\frac{m_{0i}\pi_i^2}{\brc{1-\brc{1-\rho}\pi_i}^2}-\frac{m_{2i}\brc{1-\pi_i}^2}{\brc{\rho\brc{1-\pi_i}+\pi_i}^2}\right]+\frac{m_{1+}}{\brc{1-\rho}^2}, 
\end{equation}
are used to update $\hat{\rho}^{(t+1)}$ in the $\brc{t+1}$-th iteration described in the iteration procedure.

\subsection{Dallal's Model}
Under Dallal's model the conditional probability is assumed to be a constant \cite{Dallal_1988}, i.e., 
\begin{equation}
    Pr\brc{Z_{ijk}=1|Z_{ij,3-k}=1}=\gamma. 
\end{equation}
The resulting intra-subject correlation is $\corr\brc{Z_{ijk},Z_{ij,3-k}}=\brc{\gamma-\pi_i}/\brc{1-\pi_i}$. Being bounded by the region of the probabilities and correlation, it can be shown that the region of $\gamma$ is $0\le\gamma\le1$ if $a\le1/2$; $2-1/a\le\gamma\le1$ if $a>1/2$, where $a=\underset{i=1,\ldots,g}{\max}\left\{\pi_i\right\}$. 

Taking the nuisance parameter $\kappa=\gamma$, the normal equations read  
\begin{subequations}
    \begin{equation}
        \frac{\partial l\brc{\bbeta}}{\partial\pi_i}=
        -\frac{\brc{2-\gamma}m_{0i}}{1-\brc{2-\gamma}\pi_i}+\frac{m_{1i}}{\pi_i}+\frac{m_{2i}}{\pi_i}-\frac{n_{0i}}{1-\pi_i}+\frac{n_{1i}}{\pi_i}=0, 
        \label{eq:norm:mle1:dallal:1}
    \end{equation}
    \begin{equation}
        \frac{\partial l\brc{\bbeta}}{\partial\gamma}=
        \sum_{i=1}^g\frac{m_{0i}\pi_i}{1-\brc{2-\gamma}\pi_i}-\frac{m_{1+}}{1-\gamma}+\frac{m_{2+}}{\gamma}=0, 
        \label{eq:norm:mle1:dallal:2}
    \end{equation}
\end{subequations}
where further reduction on Eq. (\ref{eq:norm:mle1:dallal:1}) gives rise to a quadratic equation below 
\begin{equation}
    a_i\pi_i^2+b_i\pi_i+c_i=0, 
    \label{eq:quadratic:pi:dallal}
\end{equation}
with the coefficients 
\begin{align*}
    &a_i=\brc{2-\gamma}\brc{m_{+i}+n_{+i}}, \\
    &b_i=-\brc{2-\gamma}m_{0i}-\brc{3-\gamma}m_{1i}-\brc{3-\gamma}m_{2i}-n_{0i}-\brc{3-\gamma}n_{1i}, \\
    &c_i=m_{1i}+m_{2i}+n_{1i}.
\end{align*}
The smaller root of the quadratic equation (\ref{eq:quadratic:pi:dallal}) leads to the maximum of the log-likelihood and thus is used at the \textbf{Step 1} in the iteration procedure. The first derivative (see middle expression in (\ref{eq:norm:mle1:dallal:2})) and the second derivative of the log-likelihood w.r.t. $\gamma$
\begin{equation}
    \frac{\partial l^2\brc{\bbeta}}{\partial\gamma^2}=-\sum_{i=1}^g\frac{m_{0i}\pi_i^2}{\brc{1-\brc{2-\gamma}\pi_i}^2}-\frac{m_{1+}}{\brc{1-\gamma}^2}-\frac{m_{2+}}{\gamma^2}, 
\end{equation}
are used to update $\hat{\gamma}^{(t+1)}$ in the $\brc{t+1}$-th iteration.

\subsection{Clayton Copula Model}
According to Sklar's theorem \cite{sklar1959fonctions} every joint cumulative density function (CDF) of a random vector can be expressed in terms of its marginal CDFs and a copula $C$. In particular, for the paired organ data, it reads 
\begin{equation}
    Pr\brc{Z_{ij1}\le z_{ij1},Z_{ij2}\le z_{ij2}}=C\brc{F_{i1}\brc{z_{ij1}},F_{i2}\brc{z_{ij2}}}. 
\end{equation}

Given $\boldsymbol{Z}_{ij}=\brc{Z_{ij1},Z_{ij2}}^T\sim \mathrm{Bernoulli}\brc{\pi_i\mathbbm{1}_2}$, the joint probabilities can be written as  
\begin{equation}
    \begin{aligned}
        p_{0i}&=Pr\brc{Z_{ij1}=0,Z_{ij2}=0}=C\brc{F_{i1}\brc{0},F_{i2}\brc{0}}=C\brc{1-\pi_i,1-\pi_i}, \\
        p_{1i}&=\sum_{k=1}^2Pr\brc{Z_{ijk}=1,Z_{ij,3-k}=0}
        =\sum_{k=1}^2\left[Pr\brc{Z_{ijk}=0}-Pr\brc{Z_{ijk}=0,Z_{ij,3-k}=0}\right] \\
        &=2\left[1-\pi_i-C\brc{1-\pi_i,1-\pi_i}\right] \\
        p_{2i}&=Pr\brc{Z_{ij1}=1,Z_{ij2}=1}=1-p_{0i}-p_{1i}
        =2\pi_i-1+C\brc{1-\pi_i,1-\pi_i}. 
    \end{aligned}
    \label{eq:prob_ri:copula}
\end{equation}

Comparing Eq. (\ref{eq:prob_ri:copula}) and Eq. (\ref{eq:prob_ri}), it is straightforward that 
\begin{equation}
    \corr\brc{Z_{ij1},Z_{ij2}}=\frac{C\brc{1-\pi_i,1-\pi_i}-\brc{1-\pi_i}^2}{\pi_i\brc{1-\pi_i}}. 
\end{equation}

The Clayton copula is a type of Archimedean copula which allows modeling dependence with one parameter ($\theta$) and is particularly suited for modeling lower tail dependence \cite{clayton1978model}. Liang \textit{et. al.} \cite{liang2025testing} utilize the Clayton copula to test the homogeneity of two proportions for correlated bilateral data, where the copula is defined as\footnote{The full expression of Clayton copula is $C_{\theta}\brc{u,v}=\left[\max\brc{u^{-\theta}+v^{-\theta}-1,~0}\right]^{-1/\theta}$, for $\theta\in[-1,0)\cup(0,\infty)$. When $\theta\in(0,\infty)$, the copula exhibits positive dependence and lower tail dependence. When $\theta\in(-1,0)$, it models negative dependence with vanishing lower tail dependence.}:
\begin{equation}
    C_{\theta}\brc{u,v}=\brc{u^{-\theta}+v^{-\theta}-1}^{-1/\theta}, 
    \quad 
    \theta>0, 
    \label{eq:copula}
\end{equation}
for $0\le u,v\le 1$. 

Using the copula form in (\ref{eq:copula}), the normal equations can be written as
\begin{subequations}
    \begin{equation}
    \begin{aligned}
        \frac{\partial l\brc{\bbeta}}{\partial\pi_i}=&-\frac{2m_{0i}}{\brc{1-\pi_i}\brc{2-\brc{1-\pi_i}^{\theta}}}
        +\frac{\brc{2\brc{2\brc{1-\pi_i}^{-\theta}-1}^{-\frac{1+\theta}{\theta}}\brc{1-\pi_i}^{-1-\theta}-1}m_{1i}}{1-\pi_i+\brc{2\brc{1-\pi_i}^{-\theta}-1}^{-1/\theta}} \\
        &+\frac{2\brc{\brc{2\brc{1-\pi_i}^{-\theta}-1}^{-\frac{1+\theta}{\theta}}\brc{1-\pi_i}^{-1-\theta}-1}m_{2i}}{1-2\pi_i-\brc{2\brc{1-\pi_i}^{-\theta}-1}^{-1/\theta}} 
        -\frac{n_{0i}}{1-\pi_i}+\frac{n_{1i}}{\pi_i}=0, 
    \end{aligned}
    \label{eq:norm:mle1:copula:1}
    \end{equation}
    \begin{equation}
    \begin{aligned}
        \frac{\partial l\brc{\bbeta}}{\partial\theta}=&\theta^{-2}\sum_{i=1}^g\brc{\log\brc{2\brc{1-\pi_i}^{-\theta}-1}+\frac{2\theta\log\brc{1-\pi_i}}{2-\brc{1-\pi_i}^{\theta}}}\times\left(\rule{0pt}{30pt}\right.m_{0i}+ \\
        &\frac{m_{1i}}{1-\brc{1-\pi_i}\brc{2\brc{1-\pi_i}^{-\theta}-1}^{1/\theta}}+\frac{m_{2i}}{1-\brc{1-2\pi_i}\brc{2\brc{1-\pi_i}^{-\theta}-1}^{1/\theta}}\left.\rule{0pt}{30pt}\right)
        =0. 
    \end{aligned}
    \label{eq:norm:mle1:copula:2}
    \end{equation}
\end{subequations}

Unlike in the previous models where the first normal equation ($\partial l\brc{\bbeta}/\partial\pi_i=0$) can be further reduced to a polynomial equation so that an analytic solution for $\hat{\pi_i}^{(t)}$ can be found for a given $\hat{\kappa}^{(t)}$ at the $\brc{t}$-th iteration, with the Clayton copula model, the root of Eq. (\ref{eq:norm:mle1:copula:1}) is evaluated numerically. The second derivative of the log-likelihood w.r.t. $\theta$
\begin{equation}
    \frac{\partial l^2\brc{\bbeta}}{\partial\theta^2}=\sum_{i=1}^g\sum_{k=0}^2c_{ki}m_{ki}, 
\end{equation}
along with the first derivative shown in (\ref{eq:norm:mle1:copula:2}) are used to update $\hat{\theta}^{(t+1)}$ at the $\brc{t+1}$-th iteration, where the coefficients $c_{ki}$'s ($k=0,1,2$) are shown below
\begin{align*}
    c_{0i}=&2\theta^{-3}\brc{-\log\brc{2\brc{1-\pi_i}^{-\theta}-1}+\frac{\theta\log\brc{1-\pi_i}\brc{\brc{1-\pi_i}^{\theta}\brc{\theta\log\brc{1-\pi_i}+2}-4}}{\brc{\brc{1-\pi_i}^{\theta}-2}^2}}, \\
    c_{1i}=&\frac{\brc{1-\pi_i}^{-2\theta}\brc{2\brc{1-\pi_i}^{-\theta}-1}^{-2\brc{1+\theta}/\theta}}{\theta^4\brc{1-\pi_i-\brc{2\brc{1-\pi_i}^{-\theta}-1}^{-1/\theta}}^2}\times\left(\rule{0pt}{20pt}\right. \\
    &+\brc{1-\pi_i}\brc{2\brc{1-\pi_i}^{-\theta}-1}^{1/\theta}\brc{\brc{2-\brc{1-\pi_i}^{\theta}}\log\brc{2\brc{1-\pi_i}^{-\theta}-1}+2\theta\log\brc{1-\pi_i}}^2 \\
    &-2\theta\brc{\brc{1-\pi_i}\brc{2\brc{1-\pi_i}^{-\theta}-1}^{1/\theta}-1}\bigg(\brc{2-\brc{1-\pi_i}^{\theta}}^2\log\brc{2\brc{1-\pi_i}^{-\theta}-1} \\
    &+\theta\log\brc{1-\pi_i}\brc{4-\brc{1-\pi_i}^{\theta}\brc{2+\theta\log\brc{1-\pi_i}}}\bigg)
    \left.\rule{0pt}{20pt}\right), \\
    c_{2i}=&\frac{\brc{1-\pi_i}^{-2\theta}\brc{2\brc{1-\pi_i}^{-\theta}-1}^{-2\brc{1+\theta}/\theta}}{\theta^4\brc{1-2\pi_i-\brc{2\brc{1-\pi_i}^{-\theta}-1}^{-1/\theta}}^2}\times\left(\rule{0pt}{20pt}\right. \\
    &+\brc{1-2\pi_i}\brc{2\brc{1-\pi_i}^{-\theta}-1}^{1/\theta}\brc{\brc{2-\brc{1-\pi_i}^{\theta}}\log\brc{2\brc{1-\pi_i}^{-\theta}-1}+2\theta\log\brc{1-\pi_i}}^2 \\
    &-2\theta\brc{\brc{1-2\pi_i}\brc{2\brc{1-\pi_i}^{-\theta}-1}^{1/\theta}+1}\bigg(\brc{2-\brc{1-\pi_i}^{\theta}}^2\log\brc{2\brc{1-\pi_i}^{-\theta}-1} \\
    &+\theta\log\brc{1-\pi_i}\brc{4-\brc{1-\pi_i}^{\theta}\brc{2+\theta\log\brc{1-\pi_i}}}\bigg)
    \left.\rule{0pt}{20pt}\right).
\end{align*}

\subsection{Independence Model}
The independence model assumes no correlation between the two paired organs of the same person, i.e.,  $\corr\brc{Z_{ij1},Z_{ij2}}=0$. Thus, this model is free of nuisance parameter and the MLEs of $\pi_i$ can be directly obtained by solving the normal equation: 
\begin{equation}
    \frac{\partial l\brc{\bbeta}}{\partial\pi_i}=-\frac{-2\pi_i m_{0i}+\brc{1-2\pi_i}m_{1i}+2\brc{1-\pi_i}m_{2i}-\pi_in_{0i}+\brc{1-\pi_i}n_{1i}}{\pi_i\brc{1-\pi_i}}=0, 
\end{equation}
which yields a closed-form solution: 
\begin{equation}
    \hat{\pi}_i=\frac{m_{1i}+2m_{2i}+n_{1i}}{2m_{+i}+n_{+i}}, \quad i=1,\ldots,g\,. 
\end{equation}

It should be noted that the independence model is a special case of the aforementioned models, including Rosner's model, Donner's model and Clayton copula model. It is nested within Rosner's model as $R\to1$, within Donner's model as $\rho\to0$, and within Clayton copula model as $\theta\to0^+$.

\subsection{Saturated Model}
The saturated model treats each joint probability $p_{ki}$ and the marginal probability $\pi_i$ as free parameters, subject to the constraint $\sum_{k=0}^2p_{ki}=1$. This model serves as a reference (or `full") model in the goodness-of-fit test, as it imposes no structural assumptions on the data. The log-likelihood is given by equation (\ref{eq:ll:generic}). Using the method of Lagrange multipliers, the MLEs of the parameters are 
\begin{equation}
    \hat{p}_{ki}=\frac{m_{ki}}{m_{+i}}, 
    \quad
    \hat{\pi}_i=\frac{n_{1i}}{n_{+i}}, 
\end{equation}
for $k=0,1,2$ and $i=1,\ldots,g$.

\section{Methods for Goodness-of-fit Test}
\label{sec:methods}
Three commonly used test statistics are employed to assess the goodness-of-fit of the model: the deviance (likelihood ratio), the Pearson chi-square and the adjusted chi-square tests. These are defined as follows. 

\paragraph{Deviance}
\begin{equation}
  \begin{aligned}
  G^2&=2\sum\texttt{observed}~\log\frac{\texttt{observed}}{\texttt{expected}} \\
  &=2\sum_{i=1}^g\brc{\sum_{r=0}^2m_{ri}\log\frac{m_{ri}}{m_{+i}\hat{p}_{ri}}
    +n_{0i}\log\frac{n_{0i}}{n_{+i}\brc{1-\hat{\pi}_i}}
    +n_{1i}\log\frac{n_{1i}}{n_{+i}\hat{\pi}_i}}, 
  \end{aligned}
  \label{eq:gof:deviance}
\end{equation}

\paragraph{Pearson Chi-square}
\begin{equation}
  \begin{aligned}
    X^2&=\sum\frac{\brc{\texttt{observed}-\texttt{expected}}^2}{\texttt{expected}} \\
    &=\sum_{i=1}^g\brc{\sum_{r=0}^2\frac{\brc{m_{ri}-m_{+i}\hat{p}_{ri}}^2}{m_{+i}\hat{p}_{ri}}+\frac{\brc{n_{0i}-n_{+i}\brc{1-\hat{\pi}_i}}^2}{n_{+i}\brc{1-\hat{\pi}_i}}
    +\frac{\brc{n_{1i}-n_{+i}\hat{\pi}_i}^2}{n_{+i}\hat{\pi}_i}}, 
  \end{aligned}
  \label{eq:gof:pearson_chisq}
\end{equation}

\paragraph{Adjusted Chi-square}
\begin{equation}
  \begin{aligned}
    X^2_{adj}&=\sum\frac{\brc{\left|\texttt{observed}-\texttt{expected}\right|-1/2}^2}{\texttt{expected}} \\
    &=\sum_{i=1}^g\brc{\sum_{r=0}^2\frac{\brc{\left|m_{ri}-m_{+i}\hat{p}_{ri}\right|-1/2}^2}{m_{+i}\hat{p}_{ri}}
      +\frac{\brc{\left|n_{0i}-n_{+i}\brc{1-\hat{\pi}_i}\right|-1/2}^2}{n_{+i}\brc{1-\hat{\pi}_i}}
    +\frac{\brc{\left|n_{1i}-n_{+i}\hat{\pi}_i\right|-1/2}^2}{n_{+i}\hat{\pi}_i}}. 
  \end{aligned}
  \label{eq:gof:adj_chisq}
\end{equation}
In the expressions above, $\hat{\pi}_i$ is the MLE of $\pi_i$, and $\hat{p}_{ri}$ is the MLE of $p_{ri}$, determined by $\hat{\pi_i}$ and $\hat{\kappa}$, where $\hat{\kappa}$ denotes the MLE of the nuisance parameter in the respective parametric models. All three statistics asymptotically follow a chi-square distribution under the null hypothesis, with the degrees of freedom (DOF) equal to the difference in number of free parameters between the saturated model and the parametric model under consideration (excluding the independence model which has no nuisance parameter), i.e., 
$\texttt{DOF}=3g-\brc{g+1}=2g-1$ \footnote{For the saturated model, there are three free parameters in each group, with two in the bilateral portion and one in the unilateral portion of data. For the parametric model in the presence of one nuisance parameter, there are $g+1$ parameters.} \footnote{For purely bilateral data ($n_{ri}=0,~r=0,1;~i=1,\ldots,g$), the \texttt{DOF} is $g-1$ since there are $2g$ free parameters for the saturated model.}.

%Additionally, the AIC criterion is applied to select the better model between Rosner's and Donner's models, provided that both are deemed acceptable based on the goodness-of-it tests. The AIC is defined as
%\begin{equation}
%  \text{AIC}=2k-2l\brc{\left\{\hat{p}_{0i},\hat{p}_{1i},\hat{p}_{2i},\hat{\pi}_i\right\}_{i=1}^g}, 
%\end{equation}
%where $k=g+1$ is the number of free parameters, and $l\brc{\left\{\widehat{~\cdot~}\right\}}$ is the log-likelihood in (\ref{eq:ll}) with MLEs of $p_{ri}$ and $\pi_i$ ($r=0,1,2;~i=1,\ldots,g$). 

Additionally, we consider three bootstrap methods ($B_1,B_2,B_3$) that use the statistic $G^2$ in (\ref{eq:gof:deviance}), the statistic $X^2$ in (\ref{eq:gof:pearson_chisq}), and the probability of the observed table below 
\begin{equation}
    Pr\brc{\brc{m,n}\mid H_0}=\prod_{i=1}^g\frac{m_{+i}!}{m_{0i}!m_{1i}!m_{2i}!}\hat{p}_{0i}^{m_{0i}}\hat{p}_{1i}^{m_{1i}}\hat{p}_{2i}^{m_{2i}}\cdot\frac{n_{+i}!}{n_{0i}!n_{1i}!}\brc{1-\hat{\pi}_i}^{n_{0i}}\hat{\pi}_i^{n_{1i}}, 
    \label{eq:prob:H0}
\end{equation}
respectively, to order the bootstrap samples.  
The bootstrap procedure is outlined as follows. 
\begin{enumerate}
    \item Generate $N_B$ bootstrap samples using the estimated parameters $\hat{\pi}_i$ and $\hat{p}_{ri}$ obtained from the observed data.

    \item For each bootstrap sample, re-estimate the parameters $\hat{\pi}_i$ and $\hat{p}_{ri}$, then compute the corresponding test statistics: $G^2$, $X^2$, and $\Pr\brc{\brc{m,n}\mid H_0}$. 
    
    \item For each method, compare the bootstrap statistic to the observed statistic from the original data. Count the number of bootstrap samples for which $G^2$ or $X^2$ is \textbf{greater than} the observed value, or $\Pr\brc{\brc{m,n}\mid H_0}$ is \textbf{less than} the observed probability. The null hypothesis is rejected if the proportion of such samples falls below a prespecified critical value (e.g., if the number of such samples is less than $5\%\cdot N_B$).

\end{enumerate}

\section{Simulation Study}
\label{sec:simulation}
We conduct a simulation study to assess the performance of the six methods for goodness-of-fit test proposed in Section \ref{sec:methods}, by investigating the empirical type I error and powers under Rosner's, Donner's, Dallal's and Clayton copula model introduced in Section \ref{sec:models}. 

\subsection{Empirical Type I Error (TIE)}
We consider equal sample size $\brc{m_+,n_+}=25,50,100$ for $g=2,4,8$, where $\brc{m_+,n_+}$ is a short notation for 
$$
\brc{m_+,n_+}=\brc{m_{+1},\ldots,m_{+g},n_{+1},\ldots,n_{+g}}.
$$
The true values of the parameters under null hypothesis $H_0$ are given in Tables. \ref{tbl:rr:H0} and \ref{tbl:pi:H0}. 
The procedure of computing the empirical type I error rates is outlined as follows. 
\begin{enumerate}
    \item[1.] Generate a dataset with a designed sample size and parameter configuration as one of the combinations given in Tables. \ref{tbl:rr:H0} and \ref{tbl:pi:H0}. 
    \item[2a.] Calculate the MLEs $\hat{\pi}_i$ and $\hat{p}_{ri}$ for each model and substitute them into Eqs. (\ref{eq:gof:deviance}) - (\ref{eq:gof:adj_chisq}) to obtain the three statistics ($G^2$, $X^2$ and $X^2_{adj}$). Reject $H_0$ if the statistics $>\chi^2_{1-\alpha;~2g-1}$; 
    \item[2b.] For bootstrap methods ($B_1,B_2,B_3$), use the generated dataset in \textbf{Step 1} as the observed data, and follow the bootstrap procedure described under Eq. (\ref{eq:prob:H0}) in Section \ref{sec:methods}.
    \item[3.] Replicate the above simulations for $N$ times, then the empirical type I error rate is computed as the ratio of the number of rejections to $N$, i.e.,
      $$\widehat{\mathrm{TIE}}=\frac{\mathrm{\#~of~rejections}}{N}.$$ 
    %% $$
    %% \widehat{\mathrm{TIE}}=\frac{\sum_{i=1}^NI\brc{Q_k>\chi^2_{1-\alpha;~2g-1}}}{N},
    %% $$
\end{enumerate}

In what follows, we set $N=10,000$ and $N_B=2,000$ to generate the simulation results.

\begin{table}[thpb]
    \centering
    \caption{Setups of the nuisance parameters in Rosner's ($R$), Donner's ($\rho$), Dallal's ($\gamma$), and Clayton copula ($\theta$) Models, under $H_0$.}
    \label{tbl:rr:H0}
    \begin{tblr}{
%    vlines,
    colspec={*{4}c},
    cell{1}{1}={c=4}{halign=c},
%    vline{2-4}={3-Z}{}
    }
        \toprule
        Model Parameter & & & \\
        \midrule
        $R$ &$\rho$ &$\gamma$ &$\theta$  \\
        \midrule
        1.2 &0.5 &0.3 &1.0 \\
        1.5 &0.7 &0.5 &2.0 \\
        1.8 &0.9 &0.7 &4.0 \\
        \bottomrule
    \end{tblr}
\end{table}

%\begin{table}[thpb]
%    \centering
%    \caption{Setups of the nuisance parameters in Rosner's ($R$), Donner's ($\rho$), Dallal's ($\gamma$), and Clayton copula ($\theta$) Models, under $H_0$.}
%    \label{tbl:rr:H0}
%    \begin{tblr}{
%    colspec={|c|*{3}c|}
%    }
%        \toprule
%        $R$ &1.2 &1.5 &1.8  \\
%        \midrule
%        $\rho$ &0.5 &0.7 &0.9 \\
%        \midrule
%        $\gamma$ &0.3 &0.5 &0.7 \\
%        \midrule
%        $\theta$ &1.0 &2.0 &4.0 \\
%        \bottomrule
%    \end{tblr}
%\end{table}

\begin{table}[thpb]
    \centering
    \caption{Setups of $\boldsymbol{\pi}=\brc{\pi_1,\ldots,\pi_g}$ for $g=2,4,8$ under $H_0$.}
    \label{tbl:pi:H0}
   \begin{tblr}{
    colspec={*{3}{c}},
%    vlines,
    cell{2}{1}={r=2}{valign=h},
    cell{4}{1}={r=2}{valign=h},
    cell{6}{1}={r=2}{valign=h},
    }
        \toprule
        $g$ &Case &$\boldsymbol{\pi}=\brc{\pi_1,\ldots,\pi_g}$  \\
        \midrule
        2 &I &$\brc{0.3,0.5}$ \\
            &II &$\brc{0.5,0.5}$ \\
        \midrule
        4 &III &$\brc{0.1,0.2,0.3,0.4}$ \\
            &IV &$\brc{0.2,0.2,0.4,0.4}$ \\
        \midrule
        8 &V &$\brc{0.1,0.2,0.3,0.4,0.1,0.2,0.3,0.4}$ \\
            &VI &$\brc{0.2,0.2,0.4,0.4,0.2,0.2,0.4,0.4}$ \\
        \bottomrule
    \end{tblr}
\end{table}

Tables \ref{tab:TIE:rosner} - \ref{tab:TIE:copula} summarize the empirical type I error rates across all scenarios under Rosner's, Donner's, Dallal's, and Clayton copula model, respectively. These results should be compared against the prespecified nominal level $\alpha$. Following the definition in Tang \textit{et al.} \cite{Tang_2006}, a statistical test is considered \textit{liberal} if the ratio of its associated empirical type I error rate to the nominal type I error rate exceeds $1.2$, \textit{conservative} if the ratio is below $0.8$, and \textit{robust} if the ratio is between $0.8$ and $1.2$. Thus, for a nominal level of $\alpha=0.05$, a \textit{liberal} test is associated with an empirical type I error rate greater than $0.06$, a \textit{conservative} test with a rate smaller than $0.04$, and a \textit{robust} test with a rate between $0.04$ and $0.06$. 

From these tables, it shows that the adjusted chi-square ($X^2_{adj}$) is too conservative with overly small empirical type I error rates. As a result, this test is not recommended\footnote{The $X^2_{adj}$ test may achieve adequate control of type I error rates only when the sample size is considerably large, for example, when $\brc{m_+,n_+}=2000$.}. The deviance ($G^2$) test performs consistently well for all four models, with empirical type I error rates mostly falling within the robust range ($0.04\lesssim\widehat{\mathrm{TIE}}\lesssim0.06$). For Rosner's, Dallal's and the Clayton copula model, the Pearson chi-square ($X^2$) test tends to be moderately more conservative than the deviance ($G^2$) test when the sample size is small, but its performance improves with larger samples, producing results that fall largely within the robust range. Under Donner's model, however, the $X^2$ test performs less favorably at small sample sizes, especially as the correlation increases. The three bootstrap methods ($B_1,B_2,B_3$) demonstrate good control over the type I error rate, producing predominantly robust results for all the models. However, a few liberal outcomes are observed under Donner’s model when the sample size is small.

\subsection{Powers}
The procedure for computing powers is analogous to that used for estimating empirical type I error rates, except that it employs the true parameter settings under the alternative hypothesis $H_1$, as specified in Tables \ref{tbl:rr:H1} and \ref{tbl:pi:H1}.

\begin{table}[thpb]
    \centering
    \caption{Setups of the nuisance parameters in Rosner's ($R$), Donner's ($\rho$), Dallal's ($\gamma$), and Clayton copula ($\theta$) Models, under $H_1$.}
    \label{tbl:rr:H1}
    \begin{tblr}{
%    vlines,
    colspec={*{4}c},
%    vline{2-4}={2-Z}{}
    }
        \toprule
            &$g=2$ &$g=4$ &$g=8$ \\
        \midrule
        $R$ &$\brc{1.2,1.5}$ &$\brc{1.2,1.2,1.5,1.5}$ &$\brc{1.2,1.2,1.2,1.2,1.5,1.5,1.5,1.5}$  \\
%        \midrule
        $\rho$ &$\brc{0.5,0.7}$ &$\brc{0.5,0.5,0.7,0.7}$ &$\brc{0.5,0.5,0.5,0.5,0.7,0.7,0.7,0.7}$ \\
%        \midrule
        $\gamma$ &$\brc{0.5,0.7}$ &$\brc{0.5,0.5,0.7,0.7}$ &$\brc{0.5,0.5,0.5,0.5,0.7,0.7,0.7,0.7}$ \\
%        \midrule
        $\theta$ &$\brc{2.0,4.0}$ &$\brc{2.0,2.0,4.0,4.0}$ &$\brc{2.0,2.0,2.0,2.0,4.0,4.0,4.0,4.0}$ \\
        \bottomrule
    \end{tblr}
\end{table}

\begin{table}[thpb]
    \centering
    \caption{Setups of $\boldsymbol{\pi}=\brc{\pi_1,\ldots,\pi_g}$ for $g=2,4,8$ under $H_1$.}
    \label{tbl:pi:H1}
   \begin{tblr}{
    colspec={*{3}{c}},
%    vlines,
    cell{2}{1}={r=2}{valign=h},
    cell{4}{1}={r=2}{valign=h},
    cell{6}{1}={r=2}{valign=h},
    }
        \toprule
        $g$ &Case &$\boldsymbol{\pi}=\brc{\pi_1,\ldots,\pi_g}$  \\
        \midrule
        2 &I &$\brc{0.2,0.2}$ \\
            &II &$\brc{0.2,0.4}$ \\
        \midrule
        4 &III &$\brc{0.1,0.2,0.3,0.4}$ \\
            &IV &$\brc{0.2,0.2,0.4,0.4}$ \\
        \midrule
        8 &V &$\brc{0.1,0.2,0.3,0.4,0.1,0.2,0.3,0.4}$ \\
            &VI &$\brc{0.2,0.2,0.4,0.4,0.2,0.2,0.4,0.4}$ \\
        \bottomrule
    \end{tblr}
\end{table}

The estimated powers across all scenarios under Rosner's, Donner's, Dallal's and Clayton copula model are presented in Table \ref{tab:powers}. For each model, the deviance ($G^2$) test overall yields the largest powers among the six methods. The three bootstrap methods produce comparable power estimates, generally close to those of the $G^2$ test. Additionally, the power of $B_3$ tends to increase slightly faster than that of the other two bootstrap methods as the number of groups ($g$) increases.

%%% Rosner's model
\begin{longtblr}[
    caption={The empirical type I error rates (in \%) for Rosner's model, 
      at the nominal level of $\alpha=0.05$.},
    label={tab:TIE:rosner}
 ]{
%   vlines,
   cell{2}{1}={r=18}{valign=h},
   cell{20}{1}={r=18}{valign=h},
   cell{38}{1}={r=18}{valign=h},
   cell{2}{2}={r=6}{valign=h},
   cell{8}{2}={r=6}{valign=h},
   cell{14}{2}={r=6}{valign=h},
   cell{20}{2}={r=6}{valign=h},
   cell{26}{2}={r=6}{valign=h},
   cell{32}{2}={r=6}{valign=h},
   cell{38}{2}={r=6}{valign=h},
   cell{44}{2}={r=6}{valign=h},
   cell{50}{2}={r=6}{valign=h},
   cell{2}{3}={r=3}{valign=h},
   cell{5}{3}={r=3}{valign=h},
   cell{8}{3}={r=3}{valign=h},
   cell{11}{3}={r=3}{valign=h},
   cell{14}{3}={r=3}{valign=h},
   cell{17}{3}={r=3}{valign=h},
   cell{20}{3}={r=3}{valign=h},
   cell{23}{3}={r=3}{valign=h},
   cell{26}{3}={r=3}{valign=h},
   cell{29}{3}={r=3}{valign=h},
   cell{32}{3}={r=3}{valign=h},
   cell{35}{3}={r=3}{valign=h},
   cell{38}{3}={r=3}{valign=h},
   cell{41}{3}={r=3}{valign=h},
   cell{44}{3}={r=3}{valign=h},
   cell{47}{3}={r=3}{valign=h},
   cell{50}{3}={r=3}{valign=h},
   cell{53}{3}={r=3}{valign=h},
 }
 \toprule
 $\left(m_+,n_+\right)$ &$g$ &case &$R$      &$G^2$ &$X^2$ &$X^2_{adj}$ &$B_1$ &$B_2$ &$B_3$ \\
 \midrule
 25        &2        &I        &1.2      & 4.87    & 4.52    & 1.29    & 4.70    & 4.74    & 5.13     \\ 
% \cline{4-10}
           &         &         &1.5      & 5.70    & 5.29    & 1.53    & 5.56    & 5.62    & 5.42     \\ 
% \cline{4-10}
           &         &         &1.8      & 5.22    & 4.72    & 1.38    & 4.82    & 4.90    & 4.38     \\ 
 \cline{3-10}
           &         &II       &1.2      & 5.10    & 4.71    & 1.52    & 4.58    & 4.63    & 4.93     \\ 
% \cline{4-10}
           &         &         &1.5      & 5.69    & 5.24    & 1.45    & 5.24    & 5.14    & 5.48     \\ 
% \cline{4-10}
           &         &         &1.8      & 4.99    & 4.53    & 1.49    & 4.75    & 4.75    & 3.97     \\ 
 \cline{2-10}
           &4        &III      &1.2      & 3.14    & 4.02    & 0.32    & 4.63    & 4.23    & 4.64     \\ 
% \cline{4-10}
           &         &         &1.5      & 3.44    & 3.88    & 0.30    & 5.02    & 4.47    & 4.52     \\ 
% \cline{4-10}
           &         &         &1.8      & 3.58    & 3.61    & 0.23    & 4.49    & 4.20    & 4.34     \\ 
 \cline{3-10}
           &         &IV       &1.2      & 3.77    & 3.40    & 0.42    & 4.73    & 4.44    & 4.72     \\ 
% \cline{4-10}
           &         &         &1.5      & 4.52    & 3.99    & 0.35    & 5.27    & 5.18    & 5.00     \\ 
% \cline{4-10}
           &         &         &1.8      & 4.30    & 3.67    & 0.38    & 4.59    & 4.40    & 4.64     \\ 
 \cline{2-10}
           &8        &V        &1.2      & 2.60    & 3.57    & 0.06    & 4.56    & 4.18    & 4.46     \\ 
% \cline{4-10}
           &         &         &1.5      & 3.13    & 3.48    & 0.05    & 5.12    & 4.49    & 4.94     \\ 
% \cline{4-10}
           &         &         &1.8      & 2.79    & 2.69    & 0.05    & 4.31    & 3.96    & 4.19     \\ 
 \cline{3-10}
           &         &VI       &1.2      & 3.79    & 3.45    & 0.18    & 4.93    & 4.79    & 4.95     \\ 
% \cline{4-10}
           &         &         &1.5      & 4.32    & 3.65    & 0.11    & 5.21    & 5.22    & 5.16     \\ 
% \cline{4-10}
           &         &         &1.8      & 4.43    & 3.47    & 0.18    & 5.05    & 4.87    & 5.01     \\ 
 \midrule
 50        &2        &I        &1.2      & 4.98    & 4.53    & 1.71    & 4.77    & 4.57    & 4.77     \\ 
% \cline{4-10}
           &         &         &1.5      & 4.99    & 4.71    & 2.12    & 4.65    & 4.74    & 4.97     \\ 
% \cline{4-10}
           &         &         &1.8      & 5.44    & 5.16    & 2.21    & 5.18    & 5.10    & 5.10     \\ 
 \cline{3-10}
           &         &II       &1.2      & 4.95    & 4.64    & 2.08    & 4.74    & 4.72    & 5.02     \\ 
% \cline{4-10}
           &         &         &1.5      & 5.35    & 5.18    & 2.29    & 5.14    & 5.25    & 5.38     \\ 
% \cline{4-10}
           &         &         &1.8      & 5.08    & 4.88    & 2.34    & 4.83    & 4.84    & 4.59     \\ 
 \cline{2-10}
           &4        &III      &1.2      & 4.12    & 4.49    & 0.89    & 5.20    & 5.02    & 5.22     \\ 
% \cline{4-10}
           &         &         &1.5      & 3.94    & 4.01    & 0.93    & 4.60    & 4.69    & 4.90     \\ 
% \cline{4-10}
           &         &         &1.8      & 4.13    & 4.08    & 0.73    & 4.88    & 4.70    & 4.84     \\ 
 \cline{3-10}
           &         &IV       &1.2      & 4.41    & 4.25    & 0.94    & 4.98    & 4.76    & 5.13     \\ 
% \cline{4-10}
           &         &         &1.5      & 4.89    & 4.60    & 0.99    & 5.03    & 4.99    & 5.15     \\ 
% \cline{4-10}
           &         &         &1.8      & 5.26    & 4.64    & 1.17    & 4.98    & 5.00    & 5.19     \\ 
 \cline{2-10}
           &8        &V        &1.2      & 3.39    & 3.82    & 0.25    & 4.62    & 4.45    & 4.66     \\ 
% \cline{4-10}
           &         &         &1.5      & 3.62    & 3.64    & 0.31    & 4.88    & 4.64    & 4.87     \\ 
% \cline{4-10}
           &         &         &1.8      & 3.97    & 3.88    & 0.37    & 4.98    & 4.90    & 4.71     \\ 
 \cline{3-10}
           &         &VI       &1.2      & 4.78    & 4.42    & 0.47    & 5.17    & 5.18    & 4.97     \\ 
% \cline{4-10}
           &         &         &1.5      & 4.95    & 4.36    & 0.43    & 5.04    & 5.02    & 4.74     \\ 
% \cline{4-10}
           &         &         &1.8      & 5.54    & 4.76    & 0.62    & 5.28    & 5.21    & 5.42     \\ 
 \midrule
 100       &2        &I        &1.2      & 4.95    & 4.77    & 2.70    & 4.83    & 4.76    & 4.84     \\ 
% \cline{4-10}
           &         &         &1.5      & 5.17    & 4.95    & 2.78    & 4.88    & 4.82    & 4.98     \\ 
% \cline{4-10}
           &         &         &1.8      & 5.04    & 4.92    & 2.98    & 5.04    & 4.94    & 5.07     \\ 
 \cline{3-10}
           &         &II       &1.2      & 5.09    & 4.92    & 2.82    & 5.02    & 5.05    & 5.20     \\ 
% \cline{4-10}
           &         &         &1.5      & 5.34    & 5.32    & 3.13    & 5.23    & 5.29    & 5.34     \\ 
% \cline{4-10}
           &         &         &1.8      & 5.24    & 5.12    & 2.91    & 5.22    & 5.23    & 5.31     \\ 
 \cline{2-10}
           &4        &III      &1.2      & 4.50    & 4.53    & 1.48    & 5.20    & 5.01    & 5.17     \\ 
% \cline{4-10}
           &         &         &1.5      & 4.37    & 4.33    & 1.35    & 4.89    & 4.80    & 4.93     \\ 
% \cline{4-10}
           &         &         &1.8      & 4.98    & 4.96    & 1.71    & 5.27    & 5.38    & 5.61     \\ 
 \cline{3-10}
           &         &IV       &1.2      & 5.00    & 4.49    & 1.56    & 4.84    & 4.66    & 4.78     \\ 
% \cline{4-10}
           &         &         &1.5      & 5.58    & 5.02    & 1.86    & 5.31    & 5.12    & 5.29     \\ 
% \cline{4-10}
           &         &         &1.8      & 5.57    & 5.07    & 1.71    & 5.26    & 5.18    & 5.32     \\ 
 \cline{2-10}
           &8        &V        &1.2      & 4.20    & 4.42    & 0.70    & 4.85    & 4.93    & 5.05     \\ 
% \cline{4-10}
           &         &         &1.5      & 4.37    & 4.42    & 0.81    & 4.93    & 5.00    & 4.87     \\ 
% \cline{4-10}
           &         &         &1.8      & 4.56    & 4.50    & 0.84    & 5.14    & 5.08    & 5.22     \\ 
 \cline{3-10}
           &         &VI       &1.2      & 4.95    & 4.57    & 1.04    & 4.74    & 4.69    & 4.72     \\ 
% \cline{4-10}
           &         &         &1.5      & 5.30    & 4.65    & 1.01    & 4.88    & 4.83    & 4.94     \\ 
% \cline{4-10}
           &         &         &1.8      & 5.39    & 4.84    & 1.02    & 5.00    & 4.94    & 5.10     \\ 
 \bottomrule
\end{longtblr}

%%% Donner's model
\begin{longtblr}[
    caption={The empirical type I error rates (in \%) for Donner's model,
      at the nominal level of $\alpha=0.05$.},
    label={tab:TIE:donner}
 ]{
%   vlines,
   cell{2}{1}={r=18}{valign=h},
   cell{20}{1}={r=18}{valign=h},
   cell{38}{1}={r=18}{valign=h},
   cell{2}{2}={r=6}{valign=h},
   cell{8}{2}={r=6}{valign=h},
   cell{14}{2}={r=6}{valign=h},
   cell{20}{2}={r=6}{valign=h},
   cell{26}{2}={r=6}{valign=h},
   cell{32}{2}={r=6}{valign=h},
   cell{38}{2}={r=6}{valign=h},
   cell{44}{2}={r=6}{valign=h},
   cell{50}{2}={r=6}{valign=h},
   cell{2}{3}={r=3}{valign=h},
   cell{5}{3}={r=3}{valign=h},
   cell{8}{3}={r=3}{valign=h},
   cell{11}{3}={r=3}{valign=h},
   cell{14}{3}={r=3}{valign=h},
   cell{17}{3}={r=3}{valign=h},
   cell{20}{3}={r=3}{valign=h},
   cell{23}{3}={r=3}{valign=h},
   cell{26}{3}={r=3}{valign=h},
   cell{29}{3}={r=3}{valign=h},
   cell{32}{3}={r=3}{valign=h},
   cell{35}{3}={r=3}{valign=h},
   cell{38}{3}={r=3}{valign=h},
   cell{41}{3}={r=3}{valign=h},
   cell{44}{3}={r=3}{valign=h},
   cell{47}{3}={r=3}{valign=h},
   cell{50}{3}={r=3}{valign=h},
   cell{53}{3}={r=3}{valign=h},
 }
 \toprule
 $\left(m_+,n_+\right)$ &$g$ &case &$\rho$   &$G^2$ &$X^2$ &$X^2_{adj}$ &$B_1$ &$B_2$ &$B_3$ \\
 \midrule
 25        &2        &I        &0.5      & 5.68    & 4.97    & 1.28    & 5.18    & 5.11    & 5.13     \\ 
% \cline{4-10}
           &         &         &0.7      & 5.07    & 4.32    & 1.16    & 5.39    & 5.37    & 5.25     \\ 
% \cline{4-10}
           &         &         &0.9      & 3.03    & 2.67    & 0.69    & 7.73    & 7.78    & 8.79     \\ 
 \cline{3-10}
           &         &II       &0.5      & 5.58    & 4.99    & 1.24    & 4.94    & 5.04    & 5.09     \\ 
% \cline{4-10}
           &         &         &0.7      & 5.18    & 4.54    & 1.15    & 5.23    & 5.23    & 5.32     \\ 
% \cline{4-10}
           &         &         &0.9      & 3.46    & 3.10    & 0.79    & 8.33    & 8.36    & 9.10     \\ 
 \cline{2-10}
           &4        &III      &0.5      & 3.90    & 3.26    & 0.21    & 5.07    & 5.01    & 4.96     \\ 
% \cline{4-10}
           &         &         &0.7      & 3.41    & 2.93    & 0.16    & 5.04    & 5.25    & 4.84     \\ 
% \cline{4-10}
           &         &         &0.9      & 1.62    & 1.60    & 0.15    & 5.64    & 5.60    & 7.00     \\ 
 \cline{3-10}
           &         &IV       &0.5      & 5.10    & 4.16    & 0.35    & 5.12    & 5.03    & 5.02     \\ 
% \cline{4-10}
           &         &         &0.7      & 4.65    & 3.90    & 0.39    & 5.30    & 5.40    & 4.94     \\ 
% \cline{4-10}
           &         &         &0.9      & 2.26    & 2.05    & 0.12    & 5.62    & 5.55    & 6.46     \\ 
 \cline{2-10}
           &8        &V        &0.5      & 3.52    & 2.78    & 0.07    & 4.76    & 4.90    & 4.26     \\ 
% \cline{4-10}
           &         &         &0.7      & 2.84    & 2.48    & 0.03    & 5.19    & 5.17    & 4.72     \\ 
% \cline{4-10}
           &         &         &0.9      & 0.63    & 0.65    & 0.00    & 3.92    & 4.00    & 5.07     \\ 
 \cline{3-10}
           &         &VI       &0.5      & 5.38    & 4.04    & 0.18    & 5.21    & 5.04    & 5.09     \\ 
% \cline{4-10}
           &         &         &0.7      & 4.26    & 3.54    & 0.11    & 5.10    & 5.02    & 4.80     \\ 
% \cline{4-10}
           &         &         &0.9      & 0.91    & 0.87    & 0.01    & 3.91    & 3.89    & 5.46     \\ 
 \midrule
 50        &2        &I        &0.5      & 5.27    & 4.96    & 1.95    & 4.88    & 4.81    & 5.02     \\ 
% \cline{4-10}
           &         &         &0.7      & 5.23    & 4.81    & 1.76    & 4.92    & 4.78    & 4.81     \\ 
% \cline{4-10}
           &         &         &0.9      & 3.87    & 3.41    & 1.05    & 6.07    & 6.03    & 6.17     \\ 
 \cline{3-10}
           &         &II       &0.5      & 5.39    & 5.14    & 2.17    & 5.07    & 5.09    & 5.28     \\ 
% \cline{4-10}
           &         &         &0.7      & 5.89    & 5.51    & 2.21    & 5.46    & 5.47    & 5.59     \\ 
% \cline{4-10}
           &         &         &0.9      & 4.29    & 3.87    & 1.56    & 6.06    & 6.05    & 6.18     \\ 
 \cline{2-10}
           &4        &III      &0.5      & 5.29    & 4.62    & 0.93    & 5.20    & 5.18    & 5.25     \\ 
% \cline{4-10}
           &         &         &0.7      & 5.28    & 4.53    & 0.79    & 5.30    & 5.25    & 5.12     \\ 
% \cline{4-10}
           &         &         &0.9      & 2.78    & 2.93    & 0.25    & 5.04    & 5.29    & 4.86     \\ 
 \cline{3-10}
           &         &IV       &0.5      & 5.70    & 4.93    & 1.19    & 5.06    & 5.12    & 5.23     \\ 
% \cline{4-10}
           &         &         &0.7      & 5.85    & 4.96    & 1.07    & 5.27    & 5.25    & 5.28     \\ 
% \cline{4-10}
           &         &         &0.9      & 3.23    & 2.99    & 0.46    & 4.99    & 5.07    & 4.68     \\ 
 \cline{2-10}
           &8        &V        &0.5      & 5.47    & 4.38    & 0.34    & 5.23    & 5.06    & 5.03     \\ 
% \cline{4-10}
           &         &         &0.7      & 5.16    & 4.21    & 0.35    & 5.13    & 5.20    & 4.81     \\ 
% \cline{4-10}
           &         &         &0.9      & 2.42    & 2.55    & 0.10    & 4.88    & 4.99    & 4.99     \\ 
 \cline{3-10}
           &         &VI       &0.5      & 5.89    & 4.96    & 0.63    & 4.92    & 5.01    & 5.23     \\ 
% \cline{4-10}
           &         &         &0.7      & 5.53    & 4.71    & 0.57    & 4.75    & 4.81    & 4.88     \\ 
% \cline{4-10}
           &         &         &0.9      & 2.78    & 3.08    & 0.19    & 4.79    & 4.91    & 5.16     \\ 
 \midrule
 100       &2        &I        &0.5      & 5.21    & 5.08    & 2.62    & 5.00    & 4.98    & 5.08     \\ 
% \cline{4-10}
           &         &         &0.7      & 4.88    & 4.69    & 2.36    & 4.74    & 4.75    & 4.84     \\ 
% \cline{4-10}
           &         &         &0.9      & 4.97    & 4.51    & 1.87    & 5.02    & 5.02    & 4.79     \\ 
 \cline{3-10}
           &         &II       &0.5      & 5.24    & 5.10    & 2.81    & 5.06    & 5.10    & 5.21     \\ 
% \cline{4-10}
           &         &         &0.7      & 5.30    & 5.17    & 2.60    & 5.04    & 5.01    & 5.17     \\ 
% \cline{4-10}
           &         &         &0.9      & 5.73    & 5.09    & 2.41    & 5.68    & 5.51    & 4.94     \\ 
 \cline{2-10}
           &4        &III      &0.5      & 5.31    & 4.90    & 1.66    & 5.03    & 5.01    & 5.05     \\ 
% \cline{4-10}
           &         &         &0.7      & 5.69    & 5.06    & 1.51    & 5.19    & 5.18    & 5.19     \\ 
% \cline{4-10}
           &         &         &0.9      & 4.20    & 4.12    & 1.04    & 4.80    & 4.94    & 4.55     \\ 
 \cline{3-10}
           &         &IV       &0.5      & 5.55    & 5.11    & 1.95    & 5.33    & 5.17    & 5.43     \\ 
% \cline{4-10}
           &         &         &0.7      & 5.25    & 4.85    & 1.57    & 4.86    & 4.80    & 4.99     \\ 
% \cline{4-10}
           &         &         &0.9      & 4.62    & 4.30    & 1.06    & 4.85    & 5.02    & 4.91     \\ 
 \cline{2-10}
           &8        &V        &0.5      & 5.70    & 5.08    & 1.08    & 5.03    & 5.17    & 5.27     \\ 
% \cline{4-10}
           &         &         &0.7      & 5.88    & 5.33    & 1.09    & 5.25    & 5.38    & 5.46     \\ 
% \cline{4-10}
           &         &         &0.9      & 3.92    & 3.82    & 0.35    & 4.88    & 4.95    & 4.89     \\ 
 \cline{3-10}
           &         &VI       &0.5      & 5.57    & 5.10    & 1.17    & 5.10    & 5.11    & 5.29     \\ 
% \cline{4-10}
           &         &         &0.7      & 5.34    & 4.68    & 0.92    & 4.81    & 4.66    & 4.96     \\ 
% \cline{4-10}
           &         &         &0.9      & 4.66    & 4.37    & 0.63    & 4.99    & 5.00    & 5.12     \\ 
 \bottomrule
\end{longtblr}

%%% Dallal's model
\begin{longtblr}[
    caption={The empirical type I error rates (in \%) for Dallal's model,
      at the nominal level of $\alpha=0.05$.},
    label={tab:TIE:dallal}
 ]{
%   vlines,
   cell{2}{1}={r=18}{valign=h},
   cell{20}{1}={r=18}{valign=h},
   cell{38}{1}={r=18}{valign=h},
   cell{2}{2}={r=6}{valign=h},
   cell{8}{2}={r=6}{valign=h},
   cell{14}{2}={r=6}{valign=h},
   cell{20}{2}={r=6}{valign=h},
   cell{26}{2}={r=6}{valign=h},
   cell{32}{2}={r=6}{valign=h},
   cell{38}{2}={r=6}{valign=h},
   cell{44}{2}={r=6}{valign=h},
   cell{50}{2}={r=6}{valign=h},
   cell{2}{3}={r=3}{valign=h},
   cell{5}{3}={r=3}{valign=h},
   cell{8}{3}={r=3}{valign=h},
   cell{11}{3}={r=3}{valign=h},
   cell{14}{3}={r=3}{valign=h},
   cell{17}{3}={r=3}{valign=h},
   cell{20}{3}={r=3}{valign=h},
   cell{23}{3}={r=3}{valign=h},
   cell{26}{3}={r=3}{valign=h},
   cell{29}{3}={r=3}{valign=h},
   cell{32}{3}={r=3}{valign=h},
   cell{35}{3}={r=3}{valign=h},
   cell{38}{3}={r=3}{valign=h},
   cell{41}{3}={r=3}{valign=h},
   cell{44}{3}={r=3}{valign=h},
   cell{47}{3}={r=3}{valign=h},
   cell{50}{3}={r=3}{valign=h},
   cell{53}{3}={r=3}{valign=h},
 }
 \toprule
 $\left(m_+,n_+\right)$ &$g$ &case &$\gamma$ &$G^2$ &$X^2$ &$X^2_{adj}$ &$B_1$ &$B_2$ &$B_3$ \\
 \midrule
 25        &2        &I        &0.3      & 4.83    & 4.60    & 1.15    & 4.32    & 4.62    & 4.08     \\ 
% \cline{4-10}
           &         &         &0.5      & 5.51    & 4.83    & 1.34    & 5.05    & 5.01    & 5.09     \\ 
% \cline{4-10}
           &         &         &0.7      & 5.60    & 4.91    & 1.08    & 5.07    & 4.99    & 4.89     \\ 
 \cline{3-10}
           &         &II       &0.3      & 5.27    & 4.64    & 1.30    & 4.94    & 4.97    & 4.35     \\ 
% \cline{4-10}
           &         &         &0.5      & 5.55    & 5.06    & 1.42    & 4.85    & 4.92    & 5.03     \\ 
% \cline{4-10}
           &         &         &0.7      & 5.62    & 5.12    & 1.35    & 5.00    & 4.98    & 5.41     \\ 
 \cline{2-10}
           &4        &III      &0.3      & 2.93    & 3.02    & 0.35    & 4.82    & 4.93    & 4.70     \\ 
% \cline{4-10}
           &         &         &0.5      & 3.93    & 3.56    & 0.36    & 5.58    & 5.39    & 5.12     \\ 
% \cline{4-10}
           &         &         &0.7      & 3.93    & 3.32    & 0.25    & 5.49    & 5.43    & 5.16     \\ 
 \cline{3-10}
           &         &IV       &0.3      & 4.12    & 3.78    & 0.50    & 5.40    & 5.27    & 5.41     \\ 
% \cline{4-10}
           &         &         &0.5      & 4.94    & 4.12    & 0.42    & 5.20    & 4.89    & 4.87     \\ 
% \cline{4-10}
           &         &         &0.7      & 5.29    & 4.37    & 0.31    & 5.46    & 5.25    & 5.11     \\ 
 \cline{2-10}
           &8        &V        &0.3      & 2.27    & 2.41    & 0.05    & 4.82    & 4.89    & 4.63     \\ 
% \cline{4-10}
           &         &         &0.5      & 3.08    & 2.72    & 0.10    & 5.10    & 5.09    & 4.46     \\ 
% \cline{4-10}
           &         &         &0.7      & 3.65    & 2.87    & 0.07    & 5.61    & 5.69    & 4.95     \\ 
 \cline{3-10}
           &         &VI       &0.3      & 3.50    & 3.28    & 0.16    & 5.20    & 5.04    & 5.19     \\ 
% \cline{4-10}
           &         &         &0.5      & 5.35    & 4.57    & 0.25    & 5.47    & 5.73    & 5.32     \\ 
% \cline{4-10}
           &         &         &0.7      & 5.45    & 4.22    & 0.11    & 5.37    & 5.42    & 5.14     \\ 
 \midrule
 50        &2        &I        &0.3      & 5.56    & 5.25    & 1.87    & 5.26    & 5.31    & 5.02     \\ 
% \cline{4-10}
           &         &         &0.5      & 5.46    & 5.18    & 2.08    & 5.04    & 5.22    & 5.25     \\ 
% \cline{4-10}
           &         &         &0.7      & 5.09    & 4.81    & 1.99    & 4.67    & 4.66    & 4.84     \\ 
 \cline{3-10}
           &         &II       &0.3      & 5.25    & 4.74    & 1.87    & 4.79    & 4.80    & 4.98     \\ 
% \cline{4-10}
           &         &         &0.5      & 5.56    & 5.19    & 2.30    & 5.16    & 5.11    & 5.38     \\ 
% \cline{4-10}
           &         &         &0.7      & 5.27    & 5.03    & 2.26    & 4.99    & 4.97    & 5.22     \\ 
 \cline{2-10}
           &4        &III      &0.3      & 4.10    & 3.90    & 0.77    & 4.80    & 4.70    & 4.74     \\ 
% \cline{4-10}
           &         &         &0.5      & 5.03    & 4.37    & 0.85    & 5.01    & 4.88    & 4.98     \\ 
% \cline{4-10}
           &         &         &0.7      & 5.34    & 4.52    & 0.90    & 5.18    & 5.12    & 4.82     \\ 
 \cline{3-10}
           &         &IV       &0.3      & 5.09    & 4.65    & 0.89    & 5.16    & 5.17    & 5.12     \\ 
% \cline{4-10}
           &         &         &0.5      & 5.73    & 5.03    & 1.05    & 4.99    & 4.93    & 5.20     \\ 
% \cline{4-10}
           &         &         &0.7      & 5.64    & 4.96    & 1.16    & 4.94    & 5.02    & 5.05     \\ 
 \cline{2-10}
           &8        &V        &0.3      & 3.96    & 3.72    & 0.41    & 4.95    & 4.92    & 4.79     \\ 
% \cline{4-10}
           &         &         &0.5      & 5.28    & 4.49    & 0.37    & 5.19    & 5.20    & 5.17     \\ 
% \cline{4-10}
           &         &         &0.7      & 5.39    & 4.21    & 0.39    & 5.20    & 4.94    & 4.95     \\ 
 \cline{3-10}
           &         &VI       &0.3      & 4.96    & 4.54    & 0.57    & 4.99    & 5.02    & 5.26     \\ 
% \cline{4-10}
           &         &         &0.5      & 5.99    & 5.04    & 0.57    & 5.20    & 5.21    & 5.41     \\ 
% \cline{4-10}
           &         &         &0.7      & 5.46    & 4.64    & 0.58    & 4.69    & 4.72    & 4.85     \\ 
 \midrule
 100       &2        &I        &0.3      & 5.51    & 5.33    & 2.83    & 5.23    & 5.33    & 5.32     \\ 
% \cline{4-10}
           &         &         &0.5      & 5.37    & 5.13    & 2.62    & 5.05    & 5.03    & 5.05     \\ 
% \cline{4-10}
           &         &         &0.7      & 5.16    & 5.06    & 2.90    & 5.03    & 5.04    & 5.16     \\ 
 \cline{3-10}
           &         &II       &0.3      & 5.23    & 5.05    & 2.60    & 5.07    & 5.11    & 5.20     \\ 
% \cline{4-10}
           &         &         &0.5      & 5.22    & 5.07    & 2.54    & 5.02    & 5.06    & 5.07     \\ 
% \cline{4-10}
           &         &         &0.7      & 5.03    & 4.89    & 2.79    & 4.86    & 4.87    & 4.96     \\ 
 \cline{2-10}
           &4        &III      &0.3      & 5.57    & 5.10    & 1.58    & 5.54    & 5.43    & 5.37     \\ 
% \cline{4-10}
           &         &         &0.5      & 5.29    & 4.90    & 1.55    & 4.94    & 4.94    & 4.98     \\ 
% \cline{4-10}
           &         &         &0.7      & 5.47    & 5.07    & 1.60    & 4.99    & 5.06    & 4.99     \\ 
 \cline{3-10}
           &         &IV       &0.3      & 5.34    & 4.76    & 1.57    & 4.96    & 4.83    & 4.83     \\ 
% \cline{4-10}
           &         &         &0.5      & 5.55    & 5.27    & 2.02    & 5.16    & 5.18    & 5.29     \\ 
% \cline{4-10}
           &         &         &0.7      & 5.49    & 5.25    & 1.78    & 5.22    & 5.13    & 5.36     \\ 
 \cline{2-10}
           &8        &V        &0.3      & 4.98    & 4.53    & 0.79    & 4.78    & 4.78    & 4.85     \\ 
% \cline{4-10}
           &         &         &0.5      & 5.59    & 5.07    & 1.06    & 5.03    & 5.14    & 5.15     \\ 
% \cline{4-10}
           &         &         &0.7      & 5.56    & 4.92    & 1.06    & 4.95    & 4.87    & 5.06     \\ 
 \cline{3-10}
           &         &VI       &0.3      & 5.54    & 5.07    & 1.00    & 5.10    & 5.04    & 5.05     \\ 
% \cline{4-10}
           &         &         &0.5      & 5.03    & 4.68    & 1.10    & 4.59    & 4.70    & 4.73     \\ 
% \cline{4-10}
           &         &         &0.7      & 5.59    & 5.09    & 1.19    & 5.21    & 5.18    & 5.30     \\ 
 \bottomrule
\end{longtblr}

%%% Clayton copula's model
 \begin{longtblr}[
     caption={The empirical type I error rates (in \%) for Clayton copula model,
       at the nominal level of $\alpha=0.05$.},
     label={tab:TIE:copula}
 ]{
%   vlines,
   cell{2}{1}={r=18}{valign=m},
   cell{20}{1}={r=18}{valign=m},
   cell{38}{1}={r=18}{valign=m},
   cell{2}{2}={r=6}{valign=m},
   cell{8}{2}={r=6}{valign=m},
   cell{14}{2}={r=6}{valign=m},
   cell{20}{2}={r=6}{valign=m},
   cell{26}{2}={r=6}{valign=m},
   cell{32}{2}={r=6}{valign=m},
   cell{38}{2}={r=6}{valign=m},
   cell{44}{2}={r=6}{valign=m},
   cell{50}{2}={r=6}{valign=m},
   cell{2}{3}={r=3}{valign=m},
   cell{5}{3}={r=3}{valign=m},
   cell{8}{3}={r=3}{valign=m},
   cell{11}{3}={r=3}{valign=m},
   cell{14}{3}={r=3}{valign=m},
   cell{17}{3}={r=3}{valign=m},
   cell{20}{3}={r=3}{valign=m},
   cell{23}{3}={r=3}{valign=m},
   cell{26}{3}={r=3}{valign=m},
   cell{29}{3}={r=3}{valign=m},
   cell{32}{3}={r=3}{valign=m},
   cell{35}{3}={r=3}{valign=m},
   cell{38}{3}={r=3}{valign=m},
   cell{41}{3}={r=3}{valign=m},
   cell{44}{3}={r=3}{valign=m},
   cell{47}{3}={r=3}{valign=m},
   cell{50}{3}={r=3}{valign=m},
   cell{53}{3}={r=3}{valign=m},
 }
 \toprule
 $\left(m_+,n_+\right)$ &$g$ &case &$\theta$ &$G^2$ &$X^2$ &$X^2_{adj}$ &$B_1$ &$B_2$ &$B_3$ \\
 \midrule
 25        &2        &I        &1.0      & 6.31    & 5.45    & 1.50    & 5.56    & 5.55    & 5.48     \\ 
% \cline{4-10}
           &         &         &2.0      & 6.09    & 5.41    & 1.68    & 5.53    & 5.52    & 5.48     \\ 
% \cline{4-10}
           &         &         &4.0      & 6.17    & 5.11    & 1.38    & 5.50    & 5.35    & 5.30     \\ 
 \cline{3-10}
           &         &II       &1.0      & 5.97    & 5.20    & 1.45    & 5.38    & 5.28    & 5.44     \\ 
% \cline{4-10}
           &         &         &2.0      & 6.04    & 5.37    & 1.35    & 5.36    & 5.46    & 5.37     \\ 
% \cline{4-10}
           &         &         &4.0      & 5.90    & 5.02    & 1.31    & 5.35    & 5.31    & 5.25     \\ 
 \cline{2-10}
           &4        &III      &1.0      & 4.18    & 3.38    & 0.27    & 5.28    & 5.13    & 5.23     \\ 
% \cline{4-10}
           &         &         &2.0      & 3.94    & 3.26    & 0.26    & 5.13    & 5.13    & 5.12     \\ 
% \cline{4-10}
           &         &         &4.0      & 4.16    & 3.44    & 0.28    & 5.32    & 5.14    & 4.95     \\ 
 \cline{3-10}
           &         &IV       &1.0      & 5.78    & 4.58    & 0.46    & 5.67    & 5.59    & 5.61     \\ 
% \cline{4-10}
           &         &         &2.0      & 5.72    & 4.56    & 0.46    & 5.68    & 5.55    & 5.22     \\ 
% \cline{4-10}
           &         &         &4.0      & 5.28    & 4.07    & 0.44    & 5.08    & 5.08    & 4.84     \\ 
 \cline{2-10}
           &8        &V        &1.0      & 3.55    & 2.81    & 0.09    & 4.97    & 4.83    & 5.28     \\ 
% \cline{4-10}
           &         &         &2.0      & 3.62    & 2.86    & 0.08    & 4.92    & 4.74    & 4.74     \\ 
% \cline{4-10}
           &         &         &4.0      & 3.85    & 3.16    & 0.06    & 5.35    & 5.19    & 4.80     \\ 
 \cline{3-10}
           &         &VI       &1.0      & 5.20    & 4.06    & 0.23    & 5.25    & 5.18    & 5.36     \\ 
% \cline{4-10}
           &         &         &2.0      & 5.37    & 4.12    & 0.18    & 5.40    & 5.36    & 5.05     \\ 
% \cline{4-10}
           &         &         &4.0      & 5.32    & 3.96    & 0.25    & 5.22    & 5.18    & 4.69     \\ 
 \midrule
 50        &2        &I        &1.0      & 5.52    & 5.20    & 2.22    & 5.11    & 5.23    & 5.33     \\ 
% \cline{4-10}
           &         &         &2.0      & 5.22    & 5.02    & 2.03    & 4.94    & 4.87    & 5.15     \\ 
% \cline{4-10}
           &         &         &4.0      & 5.49    & 5.17    & 2.18    & 5.06    & 5.07    & 5.24     \\ 
 \cline{3-10}
           &         &II       &1.0      & 5.63    & 5.41    & 2.37    & 5.36    & 5.43    & 5.52     \\ 
% \cline{4-10}
           &         &         &2.0      & 5.14    & 4.89    & 2.07    & 4.76    & 4.78    & 5.01     \\ 
% \cline{4-10}
           &         &         &4.0      & 5.24    & 4.93    & 1.98    & 4.90    & 4.93    & 5.09     \\ 
 \cline{2-10}
           &4        &III      &1.0      & 4.85    & 4.39    & 0.85    & 5.18    & 5.09    & 5.55     \\ 
% \cline{4-10}
           &         &         &2.0      & 4.95    & 4.50    & 0.92    & 5.30    & 5.14    & 5.17     \\ 
% \cline{4-10}
           &         &         &4.0      & 5.08    & 4.35    & 0.89    & 5.08    & 5.01    & 5.02     \\ 
 \cline{3-10}
           &         &IV       &1.0      & 5.65    & 5.00    & 1.06    & 5.33    & 5.25    & 5.39     \\ 
% \cline{4-10}
           &         &         &2.0      & 5.90    & 5.26    & 1.26    & 5.41    & 5.43    & 5.48     \\ 
% \cline{4-10}
           &         &         &4.0      & 6.34    & 5.35    & 1.28    & 5.65    & 5.41    & 5.53     \\ 
 \cline{2-10}
           &8        &V        &1.0      & 4.69    & 4.29    & 0.29    & 5.48    & 5.33    & 5.58     \\ 
% \cline{4-10}
           &         &         &2.0      & 4.49    & 3.75    & 0.29    & 4.93    & 4.74    & 4.94     \\ 
% \cline{4-10}
           &         &         &4.0      & 4.77    & 4.03    & 0.36    & 4.80    & 4.72    & 4.82     \\ 
 \cline{3-10}
           &         &VI       &1.0      & 5.46    & 4.70    & 0.56    & 5.15    & 4.92    & 5.01     \\ 
% \cline{4-10}
           &         &         &2.0      & 6.02    & 5.23    & 0.58    & 5.50    & 5.41    & 5.57     \\ 
% \cline{4-10}
           &         &         &4.0      & 5.98    & 5.04    & 0.62    & 4.98    & 5.08    & 5.19     \\ 
 \midrule
 100       &2        &I        &1.0      & 4.97    & 4.84    & 2.74    & 4.68    & 4.72    & 4.86     \\ 
% \cline{4-10}
           &         &         &2.0      & 5.43    & 5.29    & 2.82    & 5.18    & 5.26    & 5.26     \\ 
% \cline{4-10}
           &         &         &4.0      & 5.50    & 5.26    & 2.61    & 5.19    & 5.20    & 5.30     \\ 
 \cline{3-10}
           &         &II       &1.0      & 5.12    & 5.03    & 2.93    & 4.99    & 5.02    & 5.08     \\ 
% \cline{4-10}
           &         &         &2.0      & 5.41    & 5.31    & 2.92    & 5.24    & 5.24    & 5.36     \\ 
% \cline{4-10}
           &         &         &4.0      & 4.83    & 4.70    & 2.46    & 4.74    & 4.75    & 4.84     \\ 
 \cline{2-10}
           &4        &III      &1.0      & 4.35    & 4.12    & 1.48    & 4.61    & 4.52    & 4.91     \\ 
% \cline{4-10}
           &         &         &2.0      & 5.10    & 4.83    & 1.59    & 5.13    & 5.15    & 4.99     \\ 
% \cline{4-10}
           &         &         &4.0      & 5.29    & 4.87    & 1.32    & 5.05    & 5.04    & 4.96     \\ 
 \cline{3-10}
           &         &IV       &1.0      & 5.48    & 5.11    & 2.09    & 5.12    & 5.21    & 5.31     \\ 
% \cline{4-10}
           &         &         &2.0      & 5.27    & 5.05    & 1.84    & 5.02    & 5.01    & 5.11     \\ 
% \cline{4-10}
           &         &         &4.0      & 5.64    & 5.19    & 1.87    & 5.24    & 5.13    & 5.35     \\ 
 \cline{2-10}
           &8        &V        &1.0      & 4.66    & 4.40    & 0.72    & 5.00    & 4.91    & 5.09     \\ 
% \cline{4-10}
           &         &         &2.0      & 5.25    & 4.85    & 0.93    & 5.23    & 5.17    & 5.17     \\ 
% \cline{4-10}
           &         &         &4.0      & 5.71    & 5.04    & 0.90    & 5.46    & 5.33    & 5.28     \\ 
 \cline{3-10}
           &         &VI       &1.0      & 5.76    & 5.37    & 1.17    & 5.29    & 5.44    & 5.44     \\ 
% \cline{4-10}
           &         &         &2.0      & 5.42    & 4.93    & 1.24    & 4.88    & 4.96    & 5.11     \\ 
% \cline{4-10}
           &         &         &4.0      & 5.54    & 5.07    & 1.20    & 5.05    & 5.06    & 5.17     \\ 
 \bottomrule
 \end{longtblr}

%%% Powers
\begin{longtblr}[
   caption={The powers (in \%) with $(m_+,n_+)=150$ at the nominal level of $\alpha=0.05$.},
   label={tab:powers}
 ]{
%   vlines,
   cell{2}{1}={c=8}{halign=c},
   cell{3}{1}={r=2}{valign=h},
   cell{5}{1}={r=2}{valign=h},
   cell{7}{1}={r=2}{valign=h},
   cell{9}{1}={c=8}{halign=c},
   cell{10}{1}={r=2}{valign=h},
   cell{12}{1}={r=2}{valign=h},
   cell{14}{1}={r=2}{valign=h},
   cell{16}{1}={c=8}{halign=c},
   cell{17}{1}={r=2}{valign=h},
   cell{19}{1}={r=2}{valign=h},
   cell{21}{1}={r=2}{valign=h},
   cell{23}{1}={c=8}{halign=c},
   cell{24}{1}={r=2}{valign=h},
   cell{26}{1}={r=2}{valign=h},
   cell{28}{1}={r=2}{valign=h},
 }
 \toprule
 $g$ &case &$G^2$ &$X^2$ &$X^2_{adj}$ &$B_1$ &$B_2$ &$B_3$ \\
 \midrule
 \textit{Rosner's Model} & & & & & & &              \\ 
 \midrule
 2         &I        & 7.56    & 7.23    & 3.75    & 7.11    & 7.28    & 7.16     \\ 
% \cline{2-8}
           &II       &10.40    & 9.32    & 5.50    &10.01    & 9.25    & 9.25     \\ 
 \midrule
 4         &III      & 7.06    & 6.07    & 2.59    & 7.59    & 6.58    & 7.90     \\ 
% \cline{2-8}
           &IV       &10.85    & 9.30    & 4.71    &10.43    & 9.33    & 9.32     \\ 
 \midrule
 8         &V        &18.85    &18.52    & 7.18    &20.07    &19.63    &18.51     \\ 
% \cline{2-8}
           &VI       &30.07    &28.83    &15.00    &28.83    &28.99    &29.25     \\ 
 \midrule
 \textit{Donner's Model} & & & & & & &              \\ 
 \midrule
 2         &I        &27.79    &26.88    &19.19    &27.04    &26.93    &27.16     \\ 
% \cline{2-8}
           &II       &34.95    &34.64    &25.95    &34.28    &34.48    &34.32     \\ 
 \midrule
 4         &III      &39.77    &39.27    &26.24    &38.61    &39.24    &38.68     \\ 
% \cline{2-8}
           &IV       &46.83    &46.18    &33.17    &45.88    &46.23    &45.81     \\ 
 \midrule
 8         &V        &62.09    &60.43    &40.47    &60.54    &60.45    &60.96     \\ 
% \cline{2-8}
           &VI       &68.96    &67.95    &49.75    &67.92    &67.72    &68.18     \\ 
 \midrule
 \textit{Dallal's Model} & & & & & & &              \\ 
 \midrule
 2         &I        &33.88    &33.13    &23.75    &33.06    &32.98    &33.29     \\ 
% \cline{2-8}
           &II       &45.82    &45.62    &37.07    &45.16    &45.22    &45.18     \\ 
 \midrule
 4         &III      &51.95    &50.93    &36.64    &50.87    &50.93    &50.91     \\ 
% \cline{2-8}
           &IV       &64.99    &64.26    &51.83    &64.02    &64.23    &64.03     \\ 
 \midrule
 8         &V        &82.20    &81.21    &65.66    &81.14    &81.18    &81.61     \\ 
% \cline{2-8}
           &VI       &90.69    &90.30    &80.06    &90.29    &90.16    &90.56     \\ 
 \midrule
 \textit{Clayton copula Model} & & & & & & &      \\ 
 \midrule
 2         &I        &16.15    &15.66    &10.41    &15.96    &15.88    &16.12     \\ 
% \cline{2-8}
           &II       &21.69    &21.55    &15.31    &21.24    &21.43    &20.87     \\ 
 \bottomrule
 4         &III      &20.16    &19.06    &10.89    &19.49    &19.34    &18.47     \\ 
% \cline{2-8}
           &IV       &28.20    &27.63    &17.83    &27.40    &27.65    &26.97     \\ 
 \bottomrule
 8         &V        &38.40    &37.07    &20.43    &37.42    &37.42    &36.95     \\ 
% \cline{2-8}
           &VI       &46.79    &45.69    &28.72    &45.67    &45.67    &46.36     \\ 
 \bottomrule
\end{longtblr}

\section{Real World Applications}
\label{sec:real-data}
Unlike the simulation study, where no model preference was considered due to the data being simulated under a specific model, selecting an appropriate model is essential for analyzing real-world data. To evaluate the performance of the six proposed methods for the goodness-of-fit test, we apply them to three real-world examples. 

Model selection is conducted among the following candidates: i) the independence model, ii) Rosner's model, iii) Donner's model, iv) Dallal's model, and v) the Clayton copula model. 
The Akaike Information Criterion (AIC) is used to identify the most suitable model among these candidates, provided they pass the corresponding goodness-of-fit tests. The AIC is defined as 
\begin{equation}
  \text{AIC}=2k-2l\brc{\left\{\hat{p}_{0i},\hat{p}_{1i},\hat{p}_{2i},\hat{\pi}_i\right\}_{i=1}^g}, 
\end{equation}
where $k=g+1$ is the number of free parameters, and $l\brc{\left\{\widehat{~\cdot~}\right\}}$ is the log-likelihood with MLEs of $p_{ri}$ and $\pi_i$ ($r=0,1,2;~i=1,\ldots,g$).

\subsection{Example 1}
A double-blind randomized clinical trial was conducted at two sites to compare the cefaclor and amoxicillin for the treatment of acute otitis media with effusion (OME) in 214 children \cite{mandel1982duration}. Table \ref{tab:OME} shows the presence or absence of OME (in terms of the number of cured ears) at 14 days in 203 children from the sample of 214 children treated with cefaclor and amoxicillin. 

Table \ref{tab:ome:gof} provides the p-values of the six methods for goodness-of-fit test, along with the AIC values for the five competing models. The independence model is excluded due to extremely small p-values across the six methods. The remaining four parametric models are considered acceptable, with all p-values exceeding $0.05$. Among them, Rosner’s model and the Clayton copula model yield the highest p-values (all $\gtrsim0.7$), suggesting better fit compared to Donner’s and Dallal’s models. The AIC for the Clayton copula model is slightly lower than that for Rosner’s model, indicating that the Clayton copula model provides the best fit for this dataset.

\begin{table}[thpb]
    \centering
    \caption{Number of cured ears at 14 days in children treated with cefaclor and amoxicillin.}
    \label{tab:OME}
    \begin{tabular}{cccc}
    \toprule
         &\multicolumn{2}{c}{Treatment} & \\
         \cline{2-3}
        \# of cured ears &Cefaclor &Amoxicillin &total \\
        \midrule
        0 &21 &13 &34 \\
        1 &9 &3 &12 \\
        2 &14 &15 &29 \\
        total &44 &31 &75 \\
        \midrule
        0 &38 &27 &65 \\
        1 &24 &39 &63 \\
        total &62 &66 &128 \\
        \bottomrule
    \end{tabular}
\end{table}

\begin{table}[thpb]
  \centering
  \caption{Example 1: p-values of the six methods for goodness-of-fit test and AICs for different models.}
  \label{tab:ome:gof}
  \begin{tabular}{l*{7}{c}}
    \toprule
    &\multicolumn{6}{c}{p-value} & \\
    \cline{2-7}
    model &$G^2$ &$X^2$ &$X^2_{adj}$ &$B_1$ &$B_2$ &$B_3$ &AIC \\
    \midrule
    Independence &0.0000 &0.0000 &0.0000 &0.0000 &0.0000 &0.0000 &367.4916 \\
    Ronser's &0.7327 &0.7367 &0.8796 &0.7475 &0.7515 &0.7355 &329.4285 \\
    Donner's &0.5283 &0.5385 &0.7553 &0.5206 &0.5286 &0.5186 &330.3617 \\
    Dallal's &0.2647 &0.2741 &0.4827 &0.2690 &0.2720 &0.2615 &332.1132 \\
    Clayton copula &0.7735 &0.7742 &0.9321 &0.7790 &0.7795 &0.7740 &329.2583 \\
    \bottomrule
  \end{tabular}
\end{table}

\subsection{Example 2}
The second example involves combined unilateral and bilateral data obtained from an observational study for 60 myopia patients undergoing Orthokeratology (Ortho-k), a non-surgical vision correction method that uses specialized contact lenses worn overnight to temporarily reshape the cornea and correct myopia \cite{liang2024homogeneity}. Myopia improvement is assessed by the axial length growth (ALG), where improvement is indicated if ALG is less than $0.3$ mm, and absent otherwise. For this analysis, a subset of 33 patients using three masked brands of Ortho-K (labeled as Q, Y, and W) is included \cite{liang2024many}. The observations on the number of improved myopic eyes by bands are summarized in Table \ref{tab:myopia}. 

Table \ref{tab:myopia:gof} presents the p-values of the six methods, along with the AICs for the five competing models. The independence model yields considerably smaller p-values compared to the other four models. In particular, its p-values from the bootstrap methods $B_1$ and $B_2$ fall below $0.05$, indicating poor model fit. The remaining four models are considered acceptable, with all associated p-values exceeding $0.05$. Among them, Rosner’s model achieves the lowest AIC value, suggesting it is the best model for this dataset. It is worth noting that the Pearson chi-square test should be interpreted with caution, as several cell counts in Table \ref{tab:myopia} are smaller than $5$, potentially affecting the test's validity.

\begin{table}[thpb]
    \centering
    \caption{Number of improved myopic eyes with 3 brands of Ortho-k treatment.}
    \label{tab:myopia}
    \begin{tabular}{ccccc}
    \toprule
         &\multicolumn{3}{c}{Brand} & \\
         \cline{2-4}
        \# of myopia improved eyes &Q &Y &W &total \\
        \midrule
        0 &2 &3 &3 &8 \\
        1 &1 &1 &4 &6 \\
        2 &7 &1 &6 &14 \\
        total &10 &5 &13 &28 \\
        \midrule
        0 &1 &1 &0 &2 \\
        1 &2 &0 &1 &3 \\
        total &3 &1 &1 &5 \\
        \bottomrule
    \end{tabular}
\end{table}

\begin{table}[thpb]
  \centering
  \caption{Example 2: p-values of the six methods for goodness-of-fit test and AICs for different models.}
  \label{tab:myopia:gof}
  \begin{tabular}{l*{7}{c}}
    \toprule
    &\multicolumn{6}{c}{p-value} & \\
    \cline{2-7}
    model &$G^2$ &$X^2$ &$X^2_{adj}$ &$B_1$ &$B_2$ &$B_3$ &AIC \\
    \midrule
    Independence &0.0840 &0.0935 &0.5526 &0.0135 &0.0185 &0.5820 &74.5698 \\
    Ronser's &0.7554 &0.8399 &0.9731 &0.4377 &0.5778 &1.0000 &67.5026 \\
    Donner's &0.7466 &0.8403 &0.9593 &0.3785 &0.5467 &1.0000 &67.5607 \\
    Dallal's &0.5841 &0.6859 &0.9151 &0.2499 &0.3540 &1.0000 &68.6260 \\
    Clayton copula &0.7439 &0.8335 &0.9851 &0.3056 &0.4167 &1.0000 &67.5782 \\
    \bottomrule
  \end{tabular}
\end{table}

\subsection{Example 3}
The third example, originally analyzed in Rosner's paper introducing the constant $R$ model \cite{Rosner_1982}, is based on data from an outpatient population of 218 persons aged 20 to 39 with retinitis pigmentosa (RP), who were seen at the Massachusetts Eye and Ear Infirmary between 1970 and 1979. The patients were classified into four types of genetic groups: (i) autosomal dominant RP (DOM), (ii) autosomal recessive RP (AR), (iii) sex-linked RP (SL), and (iv) isolate RP (ISO). In order to eliminate between-subject correlation, selected patients were from different families. The distribution of the number of effected eyes for persons in the four genetic groups is given in Table \ref{tab:RP}, where an eye was considered affected if the best corrected Snellen visual acuity (VA) was 20/50 or worse, and normal if VA was 20/40 or better. Note that this dataset contains only bilateral observations and may be viewed as a special case within the combined data framework.

The results for goodness-of-fit tests are shown in Table \ref{tab:RP:gof}. As in the previous examples, the independence model is excluded due to extremely small p-values across all methods, indicating poor fit. The remaining four models are considered acceptable, with all associated p-values greater than $0.05$. Among them, Donner's model and the Clayton copula model produce the highest p-values (all $\gtrsim0.6$), more than twice those observed for Dallal's model. Rosner's model yields p-values that are marginally greater than $0.05$. Between the two best-fitting models, Donner's model has a slightly lower AIC value, indicating it is the best model for this dataset. 

\begin{table}[thpb]
    \centering
    \caption{Number of affected eyes for persons in four genetic groups.}
    \label{tab:RP}
    \begin{tabular}{cccccc}
        \toprule
         &\multicolumn{4}{c}{Genetic Type} & \\
         \cline{2-5}
        \# of affected eyes &DOM &AR &SL &ISO &total \\
        \midrule
        0 &15 &7 &3 &67 &92 \\
        1 &6 &5 &2 &24 &37 \\
        2 &7 &9 &14 &57 &87 \\
        total &28 &21 &19 &148 &216 \\
%        \midrule
%        0 &0 &0 &0 &0 &0 \\
%        1 &0 &0 &0 &0 &0 \\
%        total &0 &0 &0 &0 &0 \\
        \bottomrule
    \end{tabular}
\end{table}

\begin{table}[thpb]
  \centering
  \caption{Example 3: p-values of the six methods for goodness-of-fit test and AICs for different models.}
  \label{tab:RP:gof}
  \begin{tabular}{l*{7}{c}}
    \toprule
    &\multicolumn{6}{c}{p-value} & \\
    \cline{2-7}
    model &$G^2$ &$X^2$ &$X^2_{adj}$ &$B_1$ &$B_2$ &$B_3$ &AIC \\
    \midrule
    Independence &0.0000 &0.0000 &0.0000 &0.0000 &0.0000 &0.0000 &537.6511 \\
    Ronser's &0.0595 &0.0797 &0.2032 &0.0625 &0.0715 &0.0885 &449.9490 \\
    Donner's &0.7355 &0.7206 &0.9030 &0.7550 &0.7295 &0.6690 &443.7967 \\
    Dallal's &0.2162 &0.2424 &0.4418 &0.2375 &0.2420 &0.2205 &446.9802 \\
    Clayton copula &0.7218 &0.7063 &0.8917 &0.7190 &0.6875 &0.6620 &443.8541 \\
    \bottomrule
  \end{tabular}
\end{table}

\section{Conclusions}
\label{sec:conclusions}
Selecting an appropriate statistical model that adequately fits the observed data is a key consideration in the analysis of paired organ data. While previous work has focused on goodness-of-fit test methods for purely bilateral data, this study extends the investigation to the combined structure of unilateral and bilateral outcomes. We consider six statistical models and evaluate six methods for conducting goodness-of-fit tests under these models. A simulation study is carried out to assess the performance of the six methods under different models by computing the empirical type I error rates and powers. Based on the simulation results, we draw several conclusions. Among the three commonly used test statistics, the deviance ($G^2$) test performs well across all four parametric models. In contrast, the Pearson chi-square ($X^2$) test depends more on models and performs less well when the sample size is small, especially under Donner’s model when intra-subject correlation is high. The adjusted chi-square ($\chi^2_{adj}$) is not recommended due to its overly conservative empirical type I error rates. On the other hand, the three bootstrap methods ($B_1,B_2,B_3$) generally maintain good control over type I error rates, although a few liberal outcomes were observed under Donner’s model with small samples. Overall, differences in performance among the six methods are more pronounced when the sample size is small and tend to diminish as the sample size increases. The practical application of these methods is illustrated through three real-world datasets from otolaryngologic and ophthalmologic studies.

It is important to note that the asymptotic distributions used for the $G^2$, $X^2$ and $X^2_{adj}$ statistics (as well as for bootstrap methods $B_1$ and $B_2$ that rely on the $G^2$ and $X^2$ statistics) are theoretically valid only under the large sample conditions. When the sample size is small, these asymptotic approximations may not hold, and alternative approaches such as exact methods or the bootstrap method $B_3$ should be considered. Exploring and developing more accurate small sample inference methods remains an interesting direction for future work.

\bibliographystyle{unsrt}
\bibliography{correlateddata}

\end{document}